  \documentclass[aps,onecolumn,nofootinbib]{revtex4-1}
\usepackage{graphicx,amsmath}\usepackage{color}
\draft 
\begin{document}
\title{Coherent states of nonlinear oscillators with position-dependent mass: the temporal stability and fractional revivals}
\author{Naila Amir}

\email{naila.amir@live.com, naila.amir@seecs.nust.edu.pk}
\affiliation{School of Electrical Engineering and Computer Sciences, National University of Sciences and
             Technology, Islamabad, Pakistan}
\author{Shahid Iqbal}
\affiliation{School of Natural Sciences, National University of Sciences and
             Technology, Islamabad, Pakistan}
\email{sic80@hotmail.com, siqbal@sns.nust.edu.pk}
%
%
%
\date{\today}

\begin{abstract}

We develop generalized coherent states for a class of nonlinear oscillators with position-dependent effective mass in the context of the Gazeau-Klauder formalism and discuss some of their properties. In order to investigate the temporal evolution we first explore the statistical properties by means of weighting distribution and the Mandel parameter. It is found that the temporal evolution of the coherent states may exhibit the phenomena of quantum revivals and fractional revivals for a particular choice of position-dependent mass oscillator.\\
\vspace{80pt}

\textbf{Keywords}. Position-dependent effective mass, nonlinear oscillators, Gazeau-Klauder coherent states, Mandel parameter, sub-Poissonian statistics, temporal stability, quantum revivals, fractional revivals.
\end{abstract}


\maketitle 

%

%
\section{Introduction}
\indent The history of coherent states goes back to early days of quantum mechanics when Erwin Schr\"{o}dinger was developing the wave mechanics. In 1926, he attempted to build quantum mechanical states manifesting dynamical behaviour close to classical dynamics. He succeeded to build such quantum mechanical states for the harmonic oscillator which minimize the uncertainty relation \cite{e.schr}. 
These quantum mechanical states remained dormant for more than three decades till Roy Glauber reformulated them in terms of the ladder operators of the harmonic oscillator. In a series of his seminal papers \cite{G.1} he expressed the coherent electromagnetic field by means of these states, so named as coherent states. His ground breaking work laid the foundation of new field of quantum optics. \\
\indent Due to a lot of applications in various areas of mathematics and physics \cite{grj07,gjp09}, the notion of coherent states has been generalized for the systems other than the harmonic oscillator \cite{a04,f,amp86,aagm,ajj,klu.0,b.g,q1,f1,r1}.
Most of these generalized coherent states were based on the algebraic structure of the pertaining system. Later on, an algebraic-independent generalization procedure was introduced by Gazeau and Klauder \cite{Klu.1,G.K} and coherent states have been constructed for a large variety of Hamiltonian systems \cite{spf,pra,ujp}.
To construct the Gazeau-Klauder (GK) coherent states for a Hamiltonian system one only need to known its energy spectrum.

The GK coherent states attracted a lot of attention due to their special features, such as, these states may exhibit the phenomena of quantum revivals and fractional revivals \cite{spf,pra,ujp}. These phenomena are very important in many areas of physics such as quantum optics \cite{jpb9a,jpb10}, coherence theory \cite{jpb14}, atomic physics \cite{jpb9}, quantum chaos \cite{si2006,siphd,buch}. The phenomena have been studied with great details during last two decades \cite{buch,rob,is16,ap}.   

In this article we present GK coherent states in the context of position-dependent effective mass (PDEM) systems and study the phenomena of quantum revivals and fractional revivals during their time evolution. 
PDEM systems have attracted a lot attention due to their wide range of applications \cite{p1,p2,p7,p9,p11,p12} and their various theoretical aspects have been studied \cite{r2,com,ai15,aibg,aiagcs,von,levy,ai14,ailo,aias,1,2,3,6,7} since last few decades. However, the notion of coherent states for such systems has only been discussed most recently \cite{com,ai15,aibg,aiagcs,m09,b09,r10,sco,sg12,sam,ags}.

\indent The organization of paper is as follows. In section 2, a self-contained review, on quantizing the PDEM systems and finding their solutions, is presented. A general construction of the GK coherent states for PDEM systems is presented in section 3. In order to illustrate our general results, GK coherent states for a class of non-linear oscillators with PDEM and their revival dynamics have been discussed in section 4. Finally we conclude our work in section 5.

\section{Quantization of PDEM systems and their solutions}
\indent In this section, we present a self-contained review on the procedure of quantizing the PDEM systems and obtaining their energy spectrum using algebraic method introduced recently in Ref. \cite{aias}. This algebraic technique is based on the idea of supersymmetry quantum mechanics (SUSY QM) \cite{aias,cooper1995supersymmetry,cooper2001supersymmetry,milanovic1999generation,plastino1999supersymmetric,gonul2002supersymmetric} and an integrability condition commonly known as shape invariance (SI) \cite{aias,gendenshtein1983derivation,b,bagchi2005deformed,samani2003shape,ganguly2007shape}.\\
\indent The classical Hamiltonian for a position-dependent effective mass system is given as
\begin{equation}\label{1.1}
H=\frac{p^{2}}{2m(x)}+V(x).
\end{equation}
While quantizing the Hamiltonian (\ref{1.1}), an ordering ambiguity arises due to the incompatible nature of the operators concerning momentum and spatially varying mass in the kinetic energy term. There are several choices  \cite{com,ai15,aibg,aiagcs,von,levy,ai14,ailo,aias} to quantize the kinetic energy term of the PDEM Hamiltonian. By using symmetric ordering of $m(\hat{x})$ and $\hat{p}$, initially introduced by L\'{e}vy-Leblond \cite{levy}, the equivalent kinetic energy operator turns out to be 
\begin{equation}
\hat{T}(x,p)=\frac{1}{2}\hat{p}\frac{1}{m(x)}\hat{p}.\nonumber
\end{equation}
As a result the quantum Hamiltonian takes the form
\begin{equation}\label{qh}
\hat{H}=
-\bigg(\frac{1}{2m(x)}\bigg)\frac{d^{2}}{dx^{2}}-\bigg(\frac{1}{2m(x)}\bigg)^{'}\frac{d}{dx}+V(x),
\end{equation}
where we have used $\hat{p}=-i d/dx$ and prime denotes the differentiation with respect to $``x"$. Once the quantum Hamiltonian is in hand, one can proceed for the solutions of the PDEM system. Traditionally the exact solutions of a quantum system are obtained by solving the corresponding Schr\"{o}dinger equation. However, there exist various other methods as well that can be more advantageous over this traditional approach. Most commonly used methods include algebraic method based on the concepts of supersymmetric quantum mechanics (SUSY QM) and shape invariance \cite{aibg,aiagcs,ailo,aias,gendenshtein1983derivation,b,cooper1995supersymmetry,cooper2001supersymmetry,milanovic1999generation,bagchi2005deformed,plastino1999supersymmetric,gonul2002supersymmetric,samani2003shape,ganguly2007shape}, point canonical transformations \cite{tezcan2007exact,mustafa2006d,quesne2009point}, potential algebras \cite{ailo,kamran1990lie,roy2005lie} and path integration which relates the constant mass Green's function to that of position-dependent mass \cite{chetouani1995green,mandal2000path}. For the present work, we will follow the  algebraic technique \cite{aibg,aiagcs,ailo,aias,gendenshtein1983derivation,b,cooper1995supersymmetry} to obtain the energy spectrum of the quantum system with PDEM.\\
\indent In order to obtain the solutions by using the algebraic formalism we need to factorize the quantum Hamiltonian introduced in Eq. (\ref{qh}). For this we introduce a pair of first order differential operators
\begin{eqnarray}\label{op}
\hat{A}(x,\alpha_{1})=\frac{1}{\sqrt{2m(x)}}\frac{d}{dx}+W(x,\alpha_{1}),~~
\hat{A^{\dagger}}(x,\alpha_{1})=\frac{-d}{dx}\bigg(\frac{1}{\sqrt{2m(x)}}\bigg)+W(x,\alpha_{1}),
\end{eqnarray}
such that 
\begin{equation}\label{c}
\hat{H}=\hat{H}_{-}(x,\alpha_{1})-E_{0},
\end{equation}
where $E_{0}$ is the ground-state energy of the $\hat{H}$ and  
\begin{eqnarray}\label{ph1}
\nonumber \hat{H}_{-}(x,\alpha_{1})&=&\hat{A}^{\dagger}(x,\alpha_{1})\hat{A}(x,\alpha_{1}),\\ 
&=&\frac{-1}{2m(x)}\frac{d^{2}}{dx^{2}}-\bigg(\frac{1}{2m(x)}\bigg)^{'}\frac{d}{dx}+V_{-}(x,\alpha_{1}).
\end{eqnarray}
Here $V_{-}(x,\alpha_{1})$ represents the corresponding potential for the Hamiltonian $\hat{H}_{-}(x,\alpha_{1})$. Likewise, the product of $\hat{A}$ and $\hat{A}^{\dag}$ in reverse order provides us with another Hamiltonian of the form 
\begin{equation}\label{ph}
\hat{H}_{+}(x,\alpha_{1})=\hat{A}(x,\alpha_{1})\hat{A}^{\dagger}(x,\alpha_{1})=\frac{-1}{2m(x)}\frac{d^{2}}{dx^{2}}-\bigg(\frac{1}{2m(x)}\bigg)^{'}\frac{d}{dx}+V_{+}(x,\alpha_{1}).
\end{equation}
In the literature of SUSY QM \cite{ailo,aias,cooper1995supersymmetry,cooper2001supersymmetry} the Hamiltonians $\hat{H}_{\pm}(x,\alpha_{1})$ are known as partner Hamiltonians and the corresponding potentials $V_{\pm}(x,\alpha_{1})$ are termed as partner potentials. Moreover, the function $W(x,\alpha_{1})$\footnote{For the sake of convenience we will suppress the $x-$dependence and $\alpha-$dependence of all the operators and will mention it explicitly whenever required.}, introduced in Eq. (\ref{op}), is commonly known as super-potential and it is related to the partner potentials $V_{\pm}(x,\alpha_{1})$, introduced in Eqs. (\ref{ph1}) and (\ref{ph}), as 
\begin{eqnarray}\label{2.3'}
 V_{-}(x,\alpha_{1})&=& W^{2}(x,\alpha_{1}) - \bigg(\frac{W(x,\alpha_{1})}{\sqrt{2m(x)}}\bigg)^{'},\\
V_{+}(x,\alpha_{1})&=&V_{-}(x,\alpha_{1})+\frac{2W^{'}(x,\alpha_{1})}{\sqrt{2m(x)}}-\bigg(\frac{1}{\sqrt{2m(x)}}\bigg)\bigg(\frac{1}{\sqrt{2m(x)}}\bigg)^{''}.
\end{eqnarray}
It is important to note that the super-potential $W(x,\alpha_{1})$ and the partner potentials $V_{\pm}(x,\alpha_{1})$ depend on a set of potential parameters $\alpha_{1}$, which represent the space independent properties of the potential such as range, strength and diffuseness \cite{b}. Furthermore, $W(x,\alpha_{1})$ is related to the ground-state wave function, $\psi_{0}(x,\alpha_{1})$ of the given system, by means of the following relation \cite{aias}
\begin{equation}\label{gs}
\psi_{0}(x) = \exp\bigg(-\int \sqrt{2 m(x)}~W(x)dx\bigg).
\end{equation}
Note that the construction of operators (\ref{op}), is based on the fact that they satisfy the condition \cite{aias,b} 
\begin{equation}\label{cg}
\hat{A}|\psi_{0}\rangle= 0.
\end{equation}
Together with the definition of $\hat{H}_{-}=\hat{A}^{\dagger}\hat{A}$, Eq. (\ref{cg}) provides us with the following condition
\begin{equation}\label{gsh-}
\hat{H}_{-}|\psi_{0}\rangle= 0,
\end{equation}
which implies that $|\psi_{0}\rangle=|\psi_{0}^{(-)}\rangle$ acts as the ground state of $\hat{H}_{-}$ with ground state energy $E_{0}^{(-)}=0$.\\
\indent It is important to note, the partner Hamiltonians $\hat{H}_{\pm}$ are isospectral, i.e., if the eigenvalues and the eigenfunctions of $\hat{H}_{-}$ are known, one can immediately solve for the eigenvalues and the eigenfunctions of Hamiltonian $\hat{H}_{+}$ \cite{aias}. In principle, SUSY QM provides a key ingredient to explore the exactly solvable systems, however, the relationships obtained by using the isospectral nature of $\hat{H}_{\pm}$, do not guarantee the solvability of either of the partner potentials $V_{\pm}(x)$. In order to circumvent this problem, an integrability condition commonly known as shape invariance (SI) \cite{ailo,aias,gendenshtein1983derivation}, is required which enable us to determine all eigenvalues and eigenfunctions of both
partners without solving their Schr\"{o}dinger equations.
The partner potentials $V_{\pm}$ defined in
Eq. (\ref{2.3'}) are said to be shape invariant if they share same shape but they differ only up to a change of parameters $\alpha$ and additive constants \cite{ailo,aias,gendenshtein1983derivation,cooper1995supersymmetry,cooper2001supersymmetry}. Mathematically, the SI condition reads as \cite{aias,b}
\begin{equation}\label{2.7'}
V_{+}(x,\alpha_{1})=V_{-}(x,\alpha_{2})+R(\alpha_{1}),
\end{equation}
which in term of the partner Hamiltonians can be rewritten as
\begin{equation}\label{2.7}
\hat{H}_{+}(x,\alpha_{1})=\hat{H}_{-}(x,\alpha_{2})+R(\alpha_{1}),
\end{equation}
where $\alpha_{2}=f(\alpha_{1})$ and $R(\alpha_{1})$ represents the remainder term independent of any dynamical variables.\\ 
\indent Since the partner potentials of $\hat{H_{\pm}}$, differ only by a constant, they share common eigenfunctions, and their eigenvalues are related by the same additive constants \cite{aias,cooper2001supersymmetry}, i.e.,
\begin{eqnarray}\label{nc}
\nonumber |\psi_{n}^{(+)}(\alpha_{n})\rangle&=&|\psi_{n}^{(-)}(\alpha_{n+1})\rangle,\\
E_{n}^{(+)}(\alpha_{n})&=&E_{n}^{(-)}(\alpha_{n+1})+R(\alpha_{n}),
\end{eqnarray}
which further implies that 
\[|\psi_{0}^{(+)}(\alpha_{1})\rangle=|\psi_{0}^{(-)}(\alpha_{2})\rangle
=|\psi_{0}(\alpha_{2})\rangle.\]
In order to obtain the excited states and corresponding eigenenergies of $\hat{H}_{-}$, we make use of the intertwining relation $\hat{H}_{-}(\alpha_{1})\hat{A}^{\dagger}(\alpha_{1})
=\hat{A}^{\dagger}(\alpha_{1})\hat{H}_{+}(\alpha_{1})$, which together with integrability condition (\ref{2.7}), provides us with
\begin{eqnarray}
\nonumber	\hat{H}_{-}(\alpha_{1})[\hat{A}^{\dagger}(\alpha_{1})|\psi_{0}(\alpha_{2})\rangle]
&=&
\hat{A}^{\dagger}(\alpha_{1})\hat{H}_{+}(\alpha_{1})|\psi_{0}(\alpha_{2})\rangle,\nonumber \\
&=&\hat{A}^{\dagger}(\alpha_{1})[\hat{H}_{-}(\alpha_{2})+R(\alpha_{1})]|\psi_{0}(\alpha_{2})\rangle, \nonumber \\
&=&R(\alpha_{1})[\hat{A}^{\dagger}(\alpha_{1})|\psi_{0}(\alpha_{2})\rangle],\nonumber
\end{eqnarray}
suggesting that $\hat{A}^{\dagger}(\alpha_{1})|\psi_{0}(\alpha_{2})\rangle$ is an excited state of $\hat{H}_{-}$, with eigenenergy $E^{(-)}_{1}=R(\alpha_{1})$. Moreover,
\begin{eqnarray}
\nonumber	\hat{H}_{+}(\alpha_{1})|\psi_{0}(\alpha_{2})\rangle,
&=&\hat{H_{-}}(\alpha_{2})|\psi_{0}(\alpha_{2})\rangle+R(\alpha_{1})|\psi_{0}(\alpha_{2})\rangle\\
&=&R(\alpha_{1})|\psi_{0}(\alpha_{2})\rangle.\nonumber
\end{eqnarray}
This means that $|\psi_{0}(\alpha_{2})\rangle$ is an eigenstate of $\hat{H}_{+}$ with energy $E_{0}^{(+)}=R(\alpha_{1})$, showing the isospectral nature of the partner Hamiltonians $\hat{H}_{\pm}$. Thus, by repeating the above process we finally arrive at
\begin{equation}\label{2.11}
E^{(-)}_{n} = \sum_{i=1}^{n}R(\alpha_{i}),~~E^{(-)}_{0}=0.
\end{equation}
By using the relation (\ref{2.11}), the energy spectrum of the Hamiltonian $\hat{H}$, come out to be \cite{aias}
\begin{equation}\label{2.11'}
E_{n}= E^{(-)}_{n} + E_{0},
\end{equation}
where $E^{(-)}_{n}$ are the eigenenergies of the $\hat{H}_{-}$ and $E_{0}$ is the ground state energy of $\hat{H}$. The corresponding eigenstates of the Hamiltonian $\hat{H}$ are given as \cite{aias}
\begin{eqnarray}\label{2.13}
|\psi_{1}(x,\alpha_{1})\rangle &=& \hat{A}^{\dagger}(x,\alpha_{1})|\psi_{0}(x,\alpha_{2})\rangle \nonumber \\
|\psi_{2}(x,\alpha_{1})\rangle &=& \hat{A}^{\dagger}(x,\alpha_{1})|\psi_{1}(x,\alpha_{2})\rangle \nonumber \\
. \nonumber \\
. \nonumber \\
. \nonumber \\
. \nonumber \\
|\psi_{n}(x,\alpha_{1})\rangle &=& \hat{A}^{\dagger}(x,\alpha_{1})|\psi_{n-1}(x,\alpha_{2})\rangle.
\end{eqnarray}
Thus, we conclude that SUSY QM along with the property of shape invariance provides us
with an excellent tool to determine the entire spectrum of solvable quantum systems through
a step-by-step algebraic procedure, without going into the details of solving the corresponding
Schr\"{o}dinger equation \cite{aias,cooper1995supersymmetry,cooper2001supersymmetry}. 
\section{GK coherent states for PDEM systems}
\indent As mentioned in the introductory section, GK coherent states \cite{Klu.1,G.K} can be constructed by knowing the eigenstates and eigenvalues of a Hamiltonian system and an explicit knowledge of the underlying algebra is not needed. Therefore, the energy eigenvalues (\ref{2.11'}) and the corresponding eigenstates (\ref{2.13}), obtained in section 2, lead us to construct the GK coherent states for the PDEM systems which are defined as
\begin{equation}\label{g5.1}
|J,\gamma\rangle = \frac{1}{\mathcal{N}(J)}\sum_{n=0}^{\infty} \frac{J^{\frac{n}{2}} e^{-i\gamma e_{n}}}{\sqrt{\rho_{n}}} |\psi_{n}\rangle, ~~~J \geq 0,~~-\infty < \gamma < \infty.
\end{equation}
Here $|\psi_{n}\rangle$ are the eigenstates and $e_{n}$ are the dimensionless eigenenergies such that $e_{n+1}>e_{n}$ with $e_{0}=0$, which can be obtained by using Eq. (\ref{2.11'}) as 
\begin{equation}\label{de}
e_{n}=\frac{E_{n}-E_{0}}{\omega(\alpha_{1})},
\end{equation}
where $\omega(\alpha_{1})$ is a constant with the dimension of energy and $E_{0}$ is the ground state energy. Moreover, $\rho_{n}$ represents the product of these dimensionless energies ${e_{n}}$, i.e.,
\begin{equation}\label{g5.2}
\rho_{n} = \prod_{i=1}^{n} e_{i}; ~~~\rho_{0}=1,
\end{equation}
and $\mathcal{N}(J)$ is the normalization constant given as
\begin{equation}\label{g5.3}
\mathcal{N}^{2}(J) = \sum_{n=0}^{\infty}\frac{J^n}{\rho_{n}},
\end{equation}
which can be chosen so that $\langle J,\gamma|J,\gamma\rangle = \mathbf{1}.$
Here the domain of the allowed values of $J,~ 0<J<R$ is determined by the radius of convergence $R=\lim_{n\rightarrow \infty} (\rho_{n})^{\frac{1}{n}}$ in the series defining $\mathcal{N}^{2}(J)$. Depending on the behaviour of $\rho_{n}$ for large $n$, the radius of convergence may be finite (any non-zero values) or infinite.\\
\indent As proposed in the original formalism \cite{Klu.1,G.K}, GK coherent states satisfy a set of properties, namely, continuity of the parameters, resolution of unity, action identity and temporal stability. In our later discussion, we will briefly discuss these properties for GK coherent states of various PDEM systems. However, in order to analyze the temporal
characteristics, we first discuss the time evolution of our constructed coherent states (\ref{g5.1}), which is given as 
\begin{equation}\label{t5.1}
U(t)|J,\gamma\rangle=\frac{1}{\mathcal{N}(J)}\sum_{n=0}^{\infty} \frac{J^{\frac{n}{2}} e^{-ie_{n}(\gamma+\omega(\alpha_{1})t) }}{\sqrt{\rho_{n}}} |\psi_{n}\rangle\equiv |J,\gamma,t\rangle, 
\end{equation}
where $U(t)=\exp{(-i\hat{H} t)}$ is time evolution operator, defined in terms of PDEM quantum Hamiltonian (\ref{qh}). 

In general coherent states may exhibit quantum recurrences at various time scales during their time evolution. For a coherent state which is sufficiently well localized around a mean excitation number $n=\langle n\rangle\equiv n_{0}$ with energy $E_{n_{0}}$ these recurrence time scales are defined \cite{spf,rob} as
\begin{equation} \label{Tr}
T_{(r)}=2\pi\left( \frac{\omega(\alpha_{1})}{r!}\frac{d^{r} e_{n}}{d n^{r}}\bigg |_{n=n_{0}}\right)^{-1},~~~~r=1,2,3..,
\end{equation}
such that  $T_{1}< T_{2}< T_{3}$, where, $T_{(1)}=T_{c}$, $T_{(2)}=T_{rev}$ and $T_{(3)}=T_{sup}$ are the classical period, the quantum revival time and the super-revival time, respectively. It is obvious from Eq. (\ref{Tr}) that the occurrence of a particular recurrence time scale, during temporal evolution of a coherent state, depends on the structure of the energy spectrum of the underlying physical system. For a system with energy spectrum which is liner in quantum number, there exists only classical periodicity and the coherent states are considered as temporally stable. Otherwise, for the systems with nonlinear energy spectrum in quantum number, the coherent states undergo a series of constructive and destructive interference due to the dephasing of the constituent eigenstates of the coherent states. As a result, their temporal evolution exhibits the phenomena of quantum revivals and fractional revivals $T_{fr}=p/q(T_{rev})$, with $p,q$ being coprime integers \cite{rob}. 

A convenient way to probe temporal characteristics of a quantum state is to calculate the autocorrelation \cite{spf,rob} defined as
\begin{equation}\label{autocs}
A(t)=\langle J,\gamma,t\vert J,\gamma \rangle=\sum_{n=0}^{\infty}\vert c_{n}\vert^{2}  e^{ie_{n}\omega(\alpha_{1})t }, 
\end{equation}
where $c_{n}=J^{n/2}e^{-i\gamma e_{n}}/\mathcal{N}(J) \sqrt{\rho_{n}}$. It is obvious from Eq.~(\ref{autocs}) that $A(t)$  is an overlap of the time-evolved coherent state on to the initial state. The modulus square of autocorrelation function takes a value between one and zero, such that for a complete overlap
it is one and for a complete dephasing it is zero. In the next section, we will discuss several example to explore the phenomena of quantum revivals and fractional revivals by means of autocorrelation. 

Furthermore, it is important to note from Eq. (\ref{autocs}) that the analysis of autocorrelation function depends on the weighting distribution $\vert c_{n}\vert^{2}$ of the coherent states. Therefore, in the remaining part of this section, we present the weighting distribution as a function of coherent state parameters. In particular, for the GK coherent states, given in (\ref{g5.1}), the probability distribution is given by
\begin{equation}\label{p30}
P_{n}=\vert c_{n}\vert^{2} = \frac{J^{n}}{N^{2}(J)\rho_{n}}.
\end{equation}
The first moment of the weighting distribution of coherent states is calculated as
\begin{eqnarray}\label{fm}
n_{0}=\langle n \rangle= \sum_{n=0}^{\infty}nP_{n},  
\end{eqnarray}
where $n_{0}$ represents the mean of the given distribution and the second moment of the probability distribution is given as
\begin{eqnarray}\label{sm}
\langle n^{2} \rangle=\sum_{n=0}^{\infty}n^{2}P_{n},  
\end{eqnarray}
which enable us to calculate the variance as
\begin{equation}\label{var}
(\Delta n)^{2}=\langle n^{2} \rangle - (\langle n \rangle)^{2}.
\end{equation}
In general, the nature of a weighting distribution is characterized by the Mandel parameter \cite{grj07,mandel}
\begin{equation}\label{p36}
Q=\frac{(\Delta n)^{2}-\langle n\rangle}{\langle n\rangle},
\end{equation}
which indicates that the weighting distribution is Poissonian in nature if $Q = 0$, sub-Poissonian if $Q < 0$ and super-Poissonian if $Q > 0$. \\
\section{Nonlinear oscillators with PDEM}
To illustrate the general formalism, presented in sections 2 and 3, we consider a class of nonlinear oscillators with position-dependent effective mass in the context of coherent states and their associated properties. These oscillators have been studied recently \cite{ailo,aias} in the context of finding solutions, ladder operators and associated algebra. It is important to remark that all these oscillators are exactly solvable and posses discrete and non-degenerate energy spectrum.Therefore, the coherent states for these oscillators can be constructed using Gazeau-Klauder approach \cite{Klu.1,G.K}.\\
\subsection{Qausi-harmonic nonlinear oscillators}
This particular class of nonlinear oscillators can be modeled by a particle with position-dependent mass trapped in quadratic potential, defined as
\begin{equation}\label{vx}
V(x)=\frac{1}{2}m(x)\alpha^{2} x^{2}.
\end{equation}
such that, by choosing various profiles of $m(x)$, we get a class of nonlinear oscillators \cite{ailo,aias}. 
Here we consider the profile of spatially varying mass as $m(x)=2[1-(\lambda x)^{2}]^{-1}$ which results in the $\lambda-$dependent non-polynomial potential of the form
\[
V(x)=\frac{\alpha^{2} x^{2}}{1-(\lambda x)^{2}}.
\]
In this particular case, the general quantum Hamiltonian, given in (\ref{qh}), takes the form 
\begin{equation}\label{qh1}
\hat{H}=\frac{1}{4}\bigg[-\bigg(1-(\lambda x)^{2}\bigg)\frac{d^{2}}{dx^{2}}+2\lambda^{2} x\frac{d}{dx}
+\frac{4\alpha^{2}x^{2}}{1-(\lambda x)^{2}}\bigg].
\end{equation}
It is important to note that the mass profile, in this case, encounters a singularity for both positive and negative values of $\lambda$ and our study of dynamics is restricted to the interior of the interval $x^{2}\leq 1/\lambda^{2}$. Thus, the quantum Hamiltonian given in Eq. (\ref{qh1}), is explicitly Hermitian in the space $L^{2}[-1/\lambda,1/\lambda]$. Also, note that for $\lambda=0$, the quantum Hamiltonian for the linear harmonic oscillator with constant unit mass is recovered.\\
\indent Using Eqs. (\ref{2.11'}) and (\ref{2.13}), the energy eigenvalues and the corresponding eigenfunctions are given as 
\begin{equation}\label{a.8}
E_{n} = \alpha \bigg[\bigg(n+\frac{1}{2}\bigg)+\frac{\mu^{2}}{4}~n(n+1)\bigg],~~~~~~~n=0,1,2,....
\end{equation}
and
\begin{equation}\label{h3.19}
\psi_{n}(\varrho)=\mathcal{N}_{n}~\mathcal{H}_{m}(\varrho,\upsilon)[1-(\mu \varrho)^{2}]^{\frac{1}{2\mu^{2}}},~~~n=0,1,2,.....,
\end{equation}
respectively (see \cite{aiagcs,aias} for detailed calculations), where $\varrho=x\sqrt{2\alpha}$ and $\mu=\lambda/\sqrt{2\alpha}$, are the dimensionless variables, $\mathcal{N}_{n}$ is the normalization constant and
\begin{equation}
\mathcal{H}_{m}(\varrho,\mu)=(-1)^{n}[1-(\mu \varrho)^{2}]^{-\frac{1}{\mu^{2}}}\frac{d^{n}}{d\varrho^{n}}
[1-(\mu \varrho)^{2}]^{\frac{1}{\mu^{2}}+n}, \nonumber
\end{equation}
are the $\mu$-dependent modified Hermite polynomials.\\
\indent In order to construct the GK coherent states for this nonlinear oscillator, we consider the energy spectrum $E_{n}$ given in Eq. (\ref{a.8}) corresponding to the Hamiltonian $\hat{H}$ given in Eq. (\ref{qh1}). By using Eq. (\ref{de}), the dimensionless form of these energy eigenvalues is given as
\begin{eqnarray}\label{g5.4}
e_{n} &=& n\big[1+\frac{\mu^{2}}{4}(n+1)\big],\nonumber \\
&=& n\big[1+\upsilon^{2}(n+1)\big],
\end{eqnarray}
where $\upsilon=\frac{\mu}{2}$, which enables us to determine the parameter $\rho_{n}$, introduced in Eq. (\ref{g5.2}), as
\begin{equation}\label{g5.5}
\rho_{n}=\frac{n!\upsilon^{2n}~\Gamma\big(2+\frac{1}{\upsilon^{2}}+n\big)}{\Gamma\big(2+\frac{1}{\upsilon^{2}}\big)},
~~~\rho_{0}=1.
\end{equation}
By making use Eq. (\ref{g5.3}), the normalization constant is calculated as
\begin{equation}\label{g5.6}
\mathcal{N}^{2}(J)=~_{0}F_{1}\bigg(2+\frac{1}{\upsilon^{2}};\frac{J}{\upsilon^{2}}\bigg),
\end{equation}
and the radius of convergence turns out to be
\begin{equation}\label{g5.7}
R=\lim_{n\rightarrow \infty} \bigg[n!\upsilon^{2n}\bigg(2+\frac{1}{\upsilon^{2}}\bigg)_{n}\bigg]^{\frac{1}{n}}
=\infty.
\end{equation}
This shows that the coherent states for the non-linear oscillator with PDEM are defined on the whole complex plane. Thus, for the Hamiltonian $\hat{H}$ introduced in Eq. (\ref{qh1}), the GK coherent states take the form as
\begin{equation}\label{g5.8}
|J,\gamma\rangle = \frac{1}{\mathcal{N}(J)}\sum_{n=0}^{\infty}
\bigg[\frac{\Gamma\big(2+\frac{1}{\upsilon^{2}}\big)}{n!~\Gamma\big(2+\frac{1}{\upsilon^{2}}+n\big)}\bigg]^{\frac{1}{2}} \bigg(\frac{J}{\upsilon^{2}}\bigg)^{\frac{n}{2}} e^{-i\gamma e_{n}}  |\psi_{n}\rangle.
\end{equation}
\indent The overlap of two GK coherent states is given by
\begin{equation}\label{g5.12}
\langle J^{'},\gamma^{'}|J,\gamma\rangle=
\frac{1}{\mathcal{N}(J)\mathcal{N}(J')}\sum_{n=0}^{\infty}
\bigg[\frac{\Gamma\big(2+\frac{1}{\upsilon^{2}}\big)}{n!~\Gamma\big(2+\frac{1}{\upsilon^{2}}+n\big)}\bigg] \frac{(JJ^{'})^{\frac{n}{2}}}{\upsilon^{2n}} e^{-i(\gamma-\gamma^{'}) e_{n}}.
\end{equation}
For $J^{'}=J$ and $\gamma^{'}=\gamma$, the above relation provides us with the normalization condition $\langle J,\gamma|J,\gamma\rangle=\mathbf{1}$. \\
\indent The GK coherent states constructed in Eq. (\ref{g5.8}), satisfy the Klauder's minimal set of conditions \cite{klu.0} that are required for any coherent state. The continuity of labeling follows from the continuity of the overlap given in Eq. (\ref{g5.12}), since
\begin{equation}\label{e5.4}
\parallel |J^{'},\gamma^{'}\rangle-|J,\gamma\rangle\parallel^{2}=
[2(1-Re \langle J^{'},\gamma^{'}|J,\gamma\rangle)]
\end{equation}
approaches zero as $(J^{'},\gamma^{'}) \rightarrow (J,\gamma)$. In order to prove the resolution of unity, we need to show that
\begin{equation}\label{g5.9}
\int |J,\gamma\rangle \langle J,\gamma| d\nu(J,\gamma)=\mathbf{1},
\end{equation}
where $d\nu(J,\gamma)=w(J)(dJ)(d\gamma)/2\pi$.
By using Eq. (\ref{g5.8}) in Eq. (\ref{g5.9}), and simplifying we finally arrive at
\begin{equation}\label{g5.10}
\int_{0}^{\infty}  \tilde{w}(J)J^{n} d\zeta = \frac{\Gamma(n+1)\Gamma(2+\frac{1}{\upsilon^{2}}+n)}
{\Gamma(2+\frac{1}{\upsilon^{2}})}\upsilon^{2n},
\end{equation}
where $\tilde{w}(J)=w(J)/N(J)$ is the weight function which can be determined by using inverse Mellin transform. By using the Mellin transform of the Meijer's $G$-function \cite{rkam}, we get the required weight function $w(J)$, as
\begin{equation}\label{g5.11}
w(J)=\frac{~_{0}F_{1}\bigg(2+\frac{1}{\upsilon^{2}};\frac{J}{\upsilon^{2}}\bigg)}
{\upsilon^{2}\Gamma(2+\frac{1}{\upsilon^{2}})}
G_{0,2}^{2,0}\bigg(
\begin{array}{c}
.... \\
0,1+\frac{1}{\upsilon^{2}}
\end{array}\bigg| \frac{J}{\upsilon^{2}}
\bigg),
\end{equation}
which satisfies the integral equation (\ref{g5.9}). 

\begin{figure*}
	\centering
	\includegraphics[width=.49\textwidth]{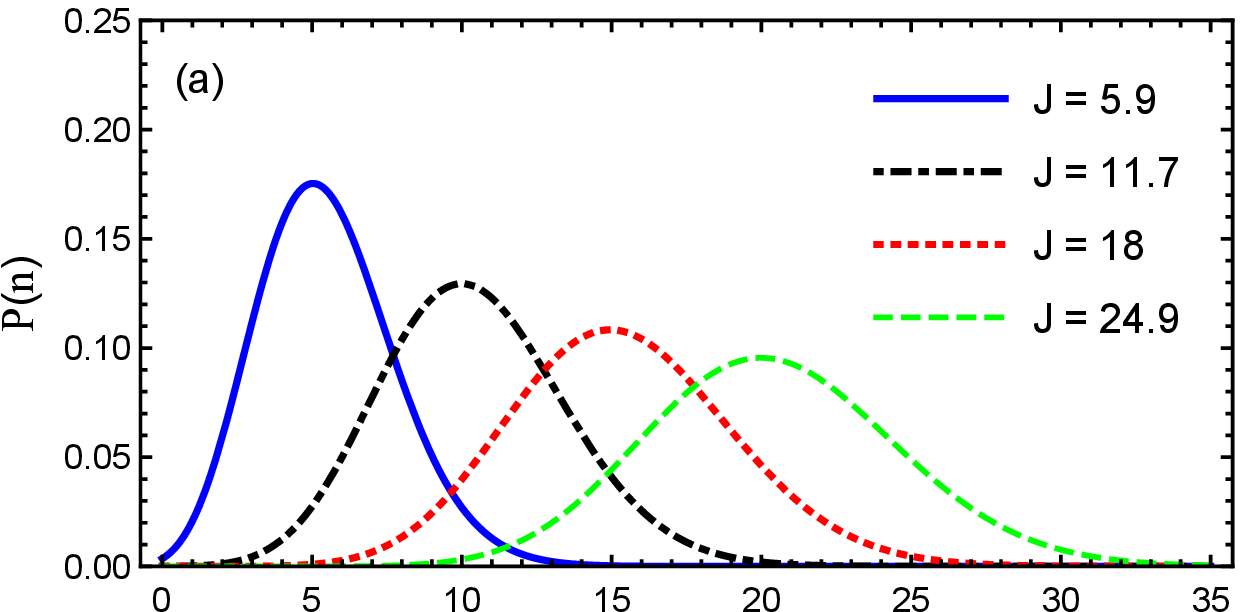}
	\includegraphics[width=.46\textwidth]{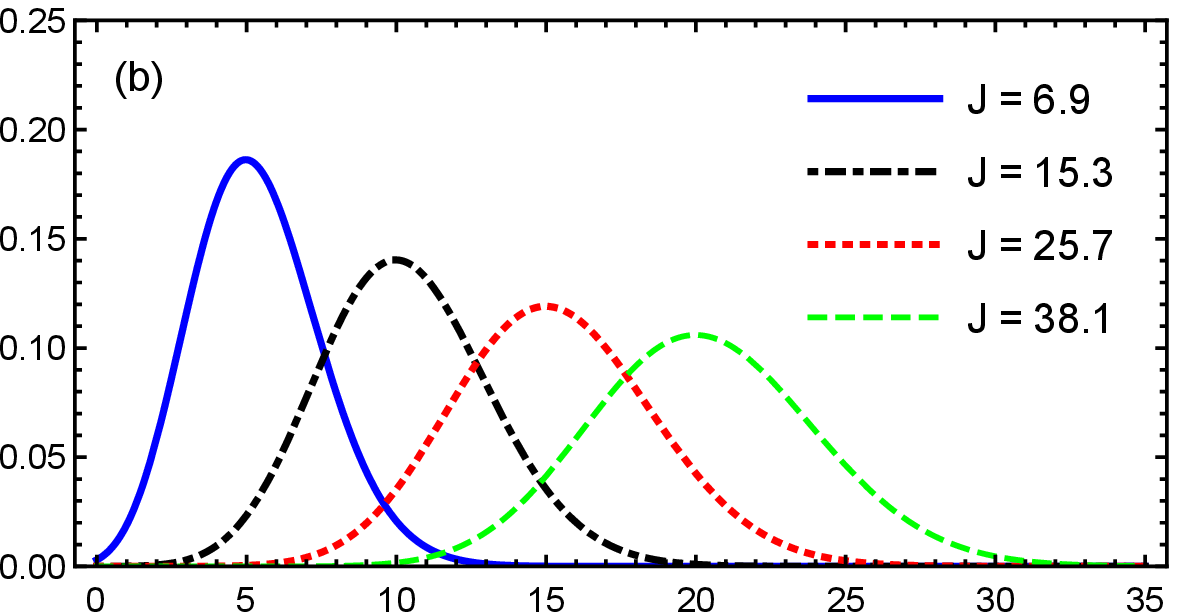}
	\includegraphics[width=.49\textwidth]{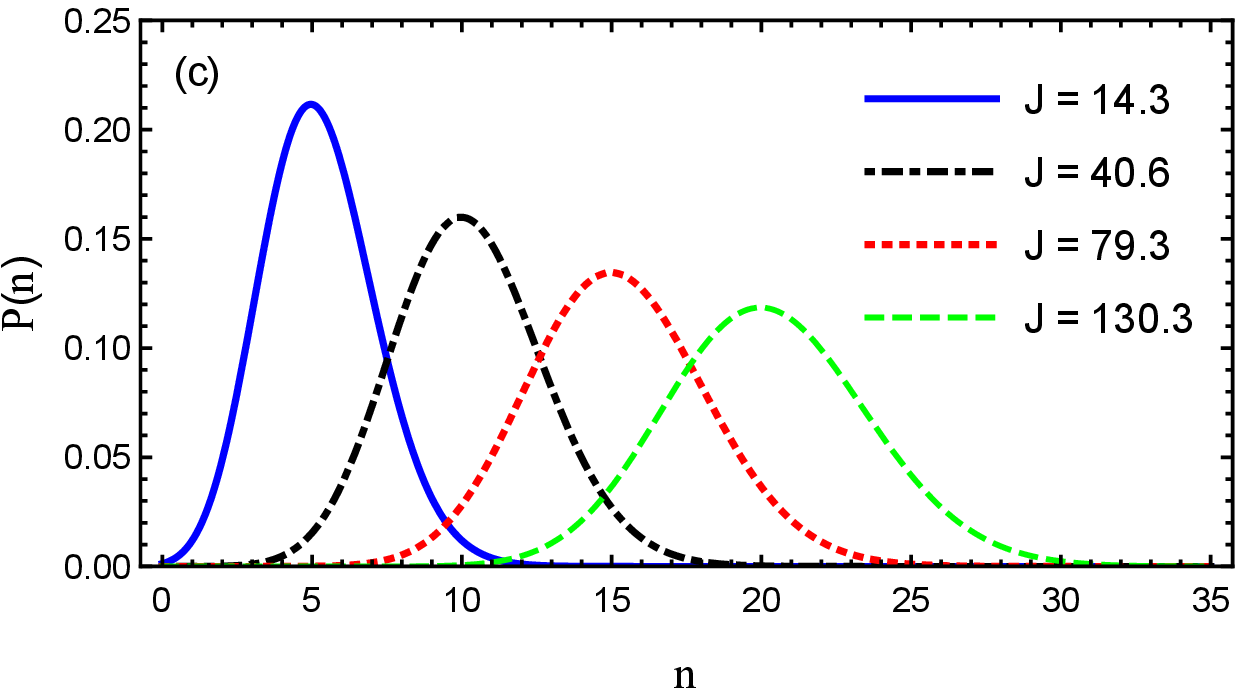}
	\includegraphics[width=.46\textwidth]{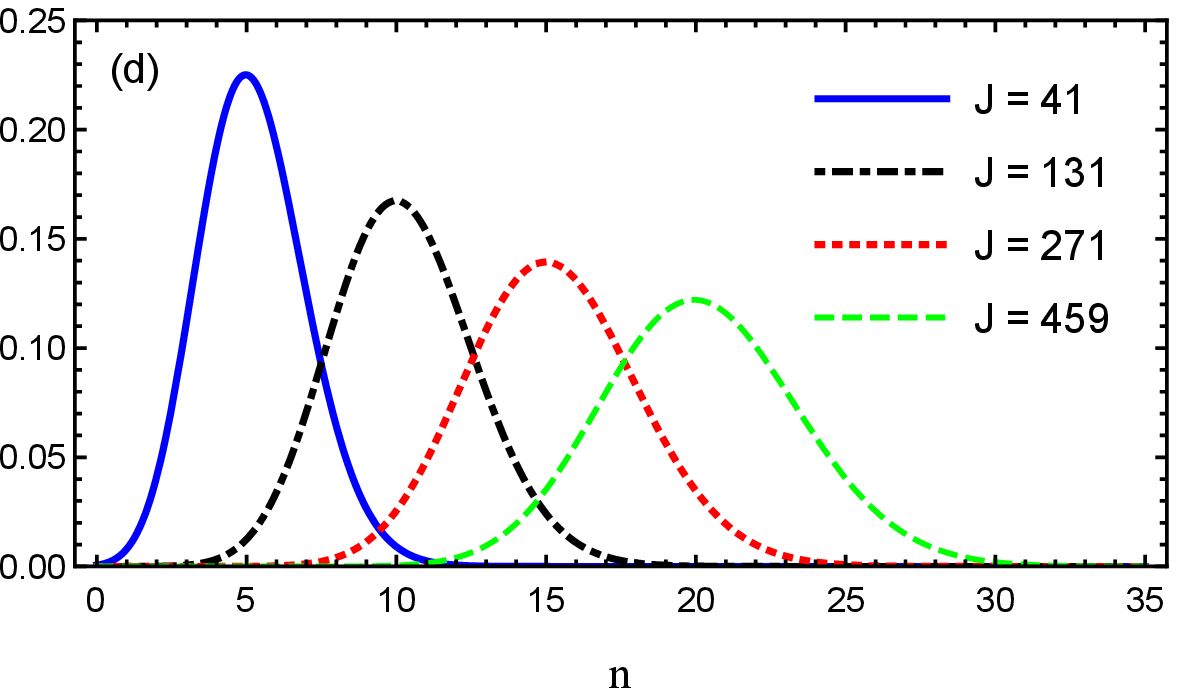}
	\caption{The weighting distribution $P(n)$ for the non-linear oscillator with PDEM as a function of quantum number $n$, for different values of the coherent state parameter $J$ and the non-linearity parameter (a) $\upsilon=0.1$, (b) $\upsilon=0.2$, (c) $\upsilon=0.5$ and (d) $\upsilon=1$.}\label{distho}
\end{figure*}

\indent Before moving to the temporal characteristics, we first examine the statistical properties introduced in the previous section. 
For the GK coherent states given in (\ref{g5.8}), the probability distribution introduced in Eq. (\ref{p30}), is given by
\begin{equation}\label{wdho}
P_{n}=\frac{1}{\mathcal{N}^{2}(J)}
\bigg[\frac{\Gamma\big(2+\frac{1}{\upsilon^{2}}\big)}{n!\Gamma\big(2+\frac{1}{\upsilon^{2}}+n\big)}\bigg]
\bigg(\frac{J}{\upsilon^{2}}\bigg)^{n}.
\end{equation}
For the sake of our later analysis, we plot the weighting distribution (\ref{wdho}) in Fig. (\ref{distho}) as a function of quantum number $n$ for different values of the non-linearity parameter $\upsilon$ and coherent state parameter $J$. In order to see the effect of the strength of position-dependence of PDEM (measured by the nonlinearity parameter $\upsilon$) on the temporal characteristics of the coherent state, we consider the values of nonlinearity parameter as $\upsilon=0.1,~0.2,~0.5,~1$ in plots (a), (b), (c) and (d) of Fig. (\ref{distho}), respectively. In each of these plots, the values of the parameter $J$ are so chosen that the corresponding coherent state is peaked at mean excitation quantum number $n_{0}=5,~10,~15,~20$. We will use these sets of parameters for the analysis of coherent state quantum revivals and fractional revivals.

\begin{figure}
	\centering
	\includegraphics[width=.49\textwidth]{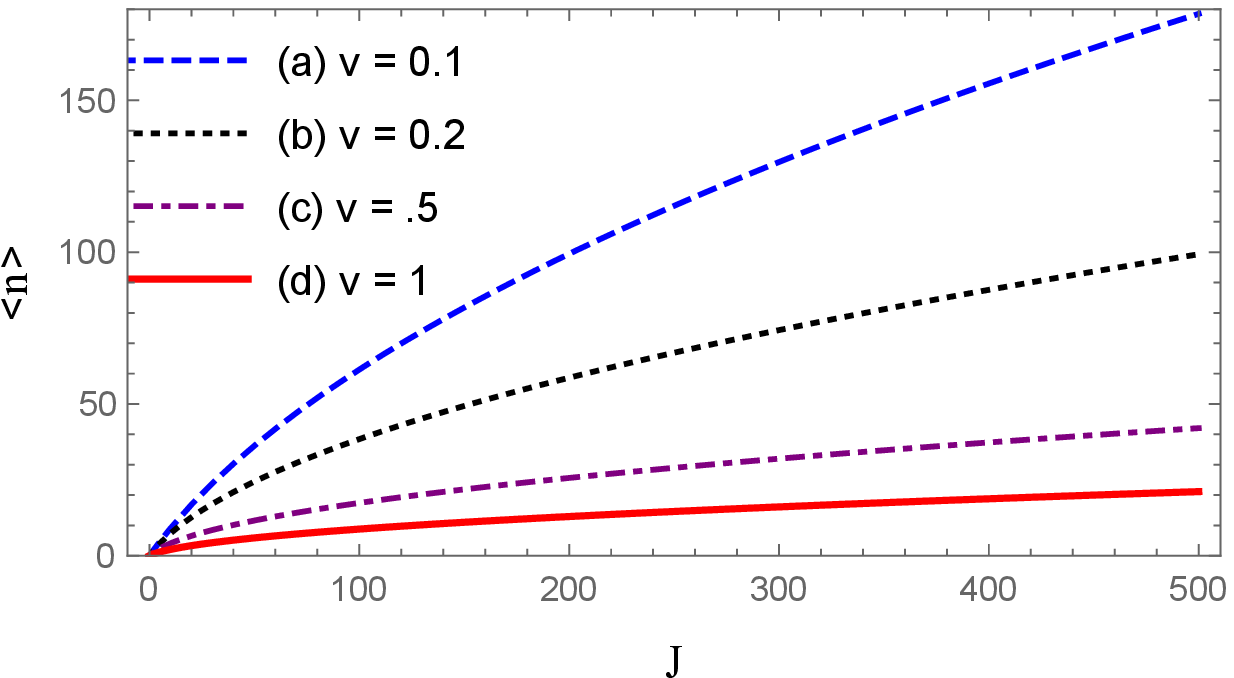}
	\includegraphics[width=.49\textwidth]{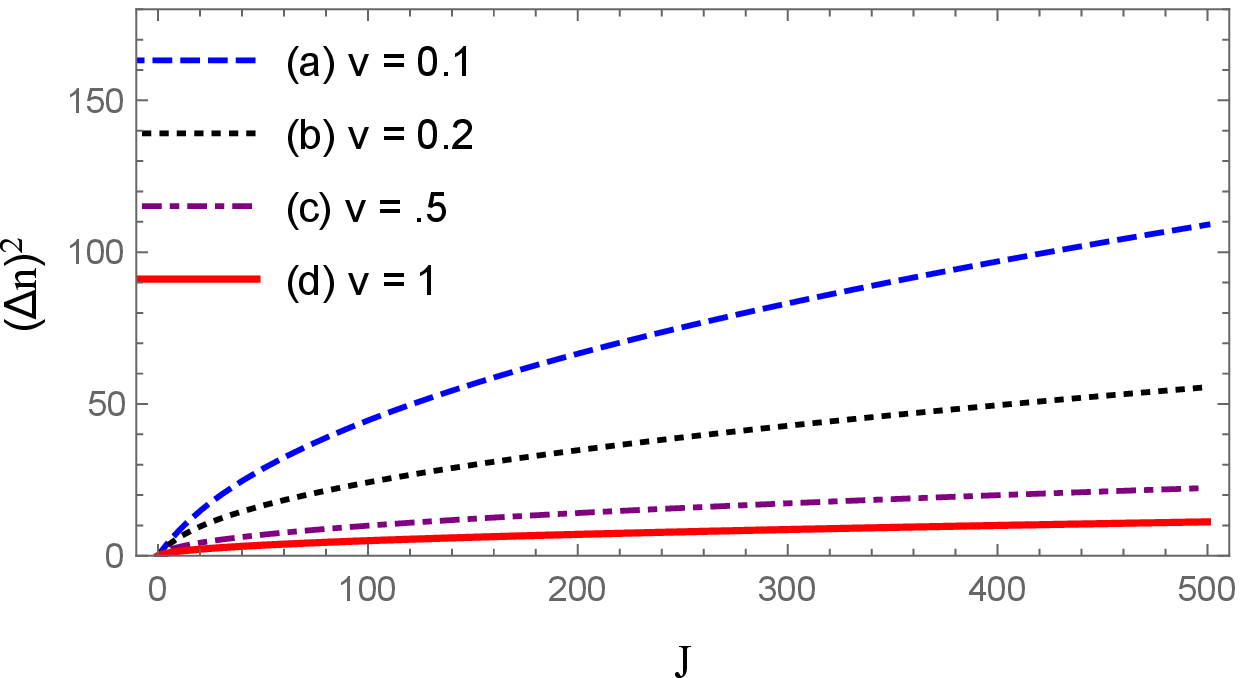}
	\caption{mean $n_{0}=\langle n\rangle$ (left) and the variance $(\Delta n)^{2}$ (right) as a function of the coherent state parameter $J$ for different values of the non-linearity parameter $(a)~\upsilon=.1$, $(b)~\upsilon=.2$, $(c)~\upsilon=.5$ and $(d)~ \upsilon=1$.} \label{mv}
\end{figure}

Moreover, in order to see an explicit dependence of the mean excitation quantum number $n_{0}$ on the system parameters ($J$ and $\upsilon$) and to see the nature of the weighting distribution, we compute the corresponding mean and the variance, respectively, defined in Eqs. (\ref{fm}) and (\ref{var}). For the weighting distribution, given in Eq. (\ref{wdho}), the means is given as
\begin{eqnarray}\label{m}
\langle n \rangle= \frac{J}{(2\upsilon^{2}+1)}
\frac{~_{0}F_{1}\bigg(3+\frac{1}{\upsilon^{2}};\frac{J}{\upsilon^{2}}\bigg)}
{~_{0}F_{1}\bigg(2+\frac{1}{\upsilon^{2}};\frac{J}{\upsilon^{2}}\bigg)},
\end{eqnarray}
and the corresponding second moment, introduced in Eq. (\ref{sm}), is given as
\begin{eqnarray}\label{m'}
\langle n^{2} \rangle= \langle n \rangle+\frac{J^{2}}{(2\upsilon^{2}+1)(3\upsilon^{2}+1)}
\frac{~_{0}F_{1}\bigg(4+\frac{1}{\upsilon^{2}};\frac{J}{\upsilon^{2}}\bigg)}
{~_{0}F_{1}\bigg(2+\frac{1}{\upsilon^{2}};\frac{J}{\upsilon^{2}}\bigg)},\nonumber
\end{eqnarray}
which leads us to calculate the variance, defined in Eq. (\ref{var}), given as
\begin{equation}\label{v}
(\Delta n)^{2}= \langle n \rangle[1-\langle n \rangle]+\frac{J^{2}}{(2\upsilon^{2}+1)(3\upsilon^{2}+1)}
\frac{~_{0}F_{1}\bigg(4+\frac{1}{\upsilon^{2}};\frac{J}{\upsilon^{2}}\bigg)}
{~_{0}F_{1}\bigg(2+\frac{1}{\upsilon^{2}};\frac{J}{\upsilon^{2}}\bigg)}.
\end{equation}
The mean and variance, obtained in Eqs. (\ref{m}) and (\ref{v}) respectively, have been plotted in Fig. (\ref{mv}) as a function of $J$ for the same set of values of $\upsilon$ as in Fig. (\ref{distho}). It is important to note from the plots in Fig. (\ref{mv}), that the mean and the variance are directly proportional to the coherent state parameter $J$.  Moreover, it is clear from the plots that for all values of the $\upsilon$, mean is grater than the variance, i.e., $\langle n \rangle > (\Delta n)^{2}$, which indicates the sub-Poissonian nature of the weighting distribution for the system under discussion.
Moreover, we can calculate Mandel parameter, introduced in Eq. (\ref{p36}), which in the present case takes the form
\[ Q=\frac{J}{(3\upsilon^{2}+1)}
\frac{~_{0}F_{1}\bigg(4+\frac{1}{\upsilon^{2}};\frac{J}{\upsilon^{2}}\bigg)}
{~_{0}F_{1}\bigg(3+\frac{1}{\upsilon^{2}};\frac{J}{\upsilon^{2}}\bigg)}-
\frac{J}{(2\upsilon^{2}+1)}
\frac{~_{0}F_{1}\bigg(3+\frac{1}{\upsilon^{2}};\frac{J}{\upsilon^{2}}\bigg)}
{~_{0}F_{1}\bigg(2+\frac{1}{\upsilon^{2}};\frac{J}{\upsilon^{2}}\bigg)}.\]
It is straightforward to show that $Q<0$ which indicates the sub-Poissonian nature of the weighting distribution for all values of $\upsilon$ except $\upsilon=0$ which recovers the Poissonian distribution of coherent states of linear oscillator. 

\begin{figure}
	\centering
	\includegraphics[width=0.49\textwidth]{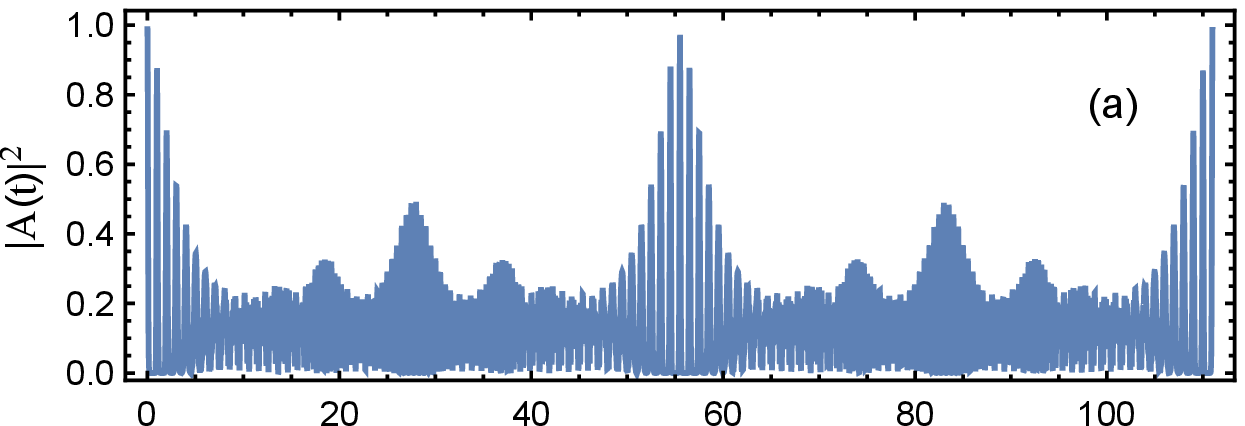}
	\includegraphics[width=0.49\textwidth]{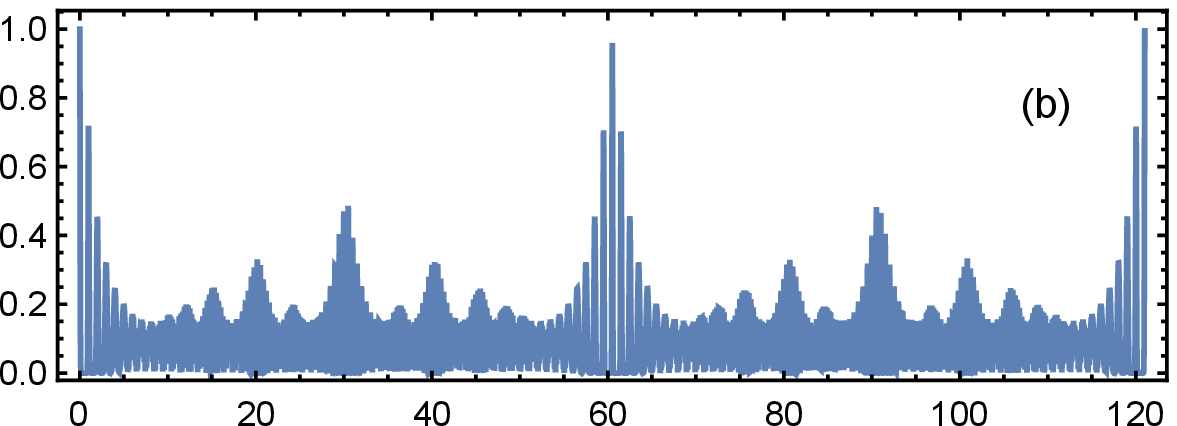}	\includegraphics[width=0.49\textwidth]{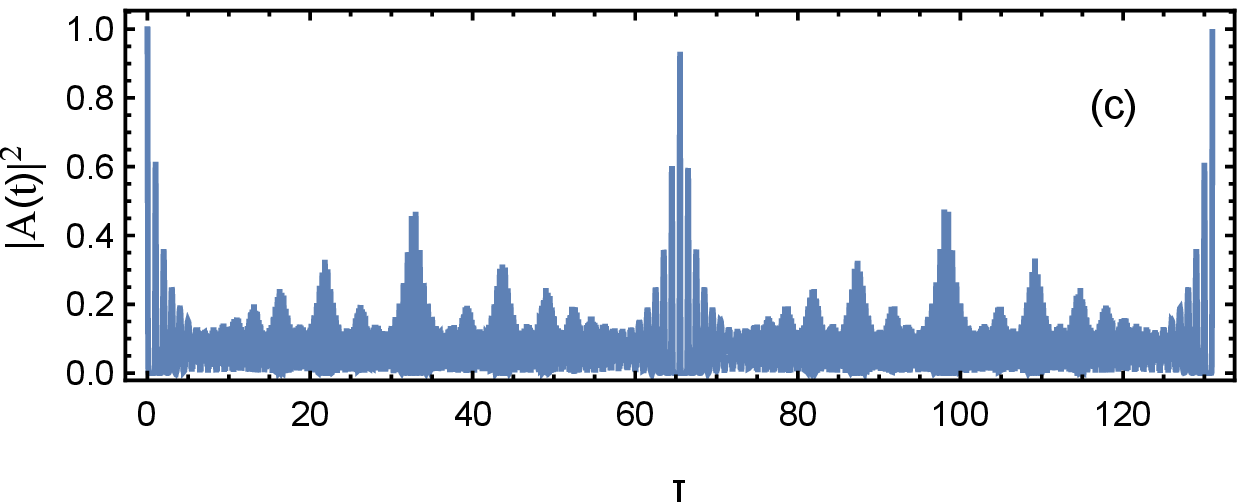}	\includegraphics[width=0.49\textwidth]{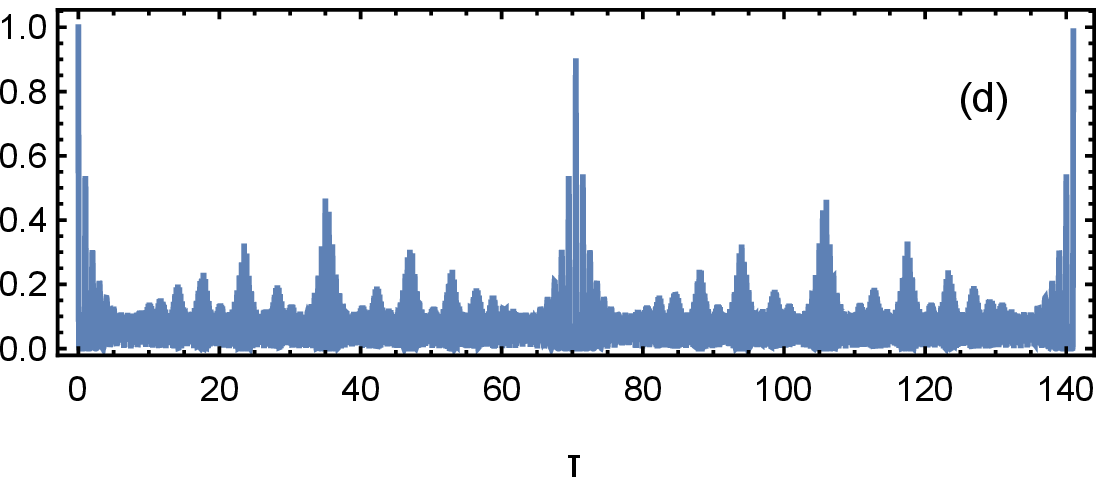}
	\caption{The modulus square of autocorrelation function $A(t)$ versus time $\tau=t/T_{cl}$ for $\upsilon=0.1$ and $(a)~n_{0}=5,~J=5.9$, $(b)~n_{0}=10,~J=11.7$, $(c)~n_{0}=15,~J=18,$ and $(d)~n_{0}=20,~J=24.9$.} \label{a.1}
\end{figure}

\begin{figure}
	\centering
	\includegraphics[width=0.49\textwidth]{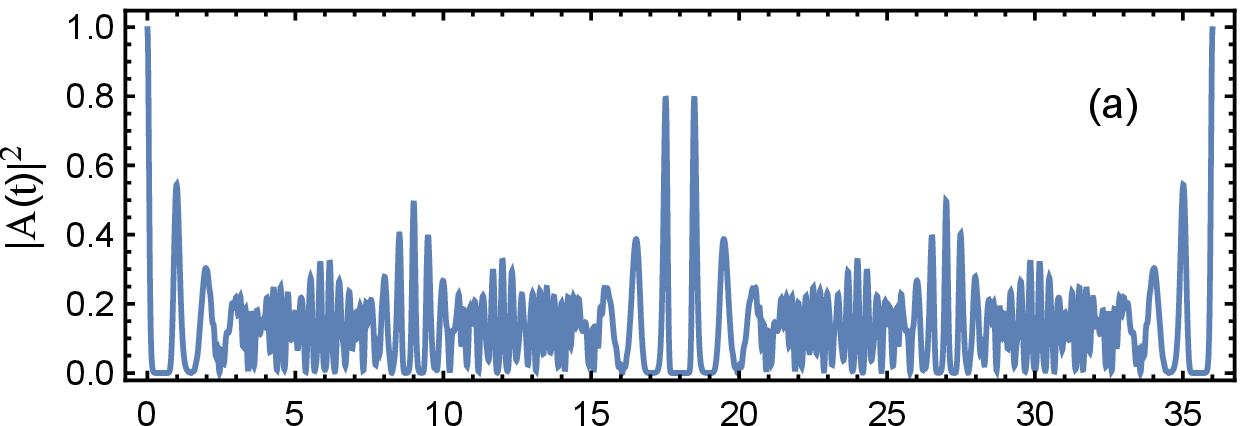}
	\includegraphics[width=0.49\textwidth]{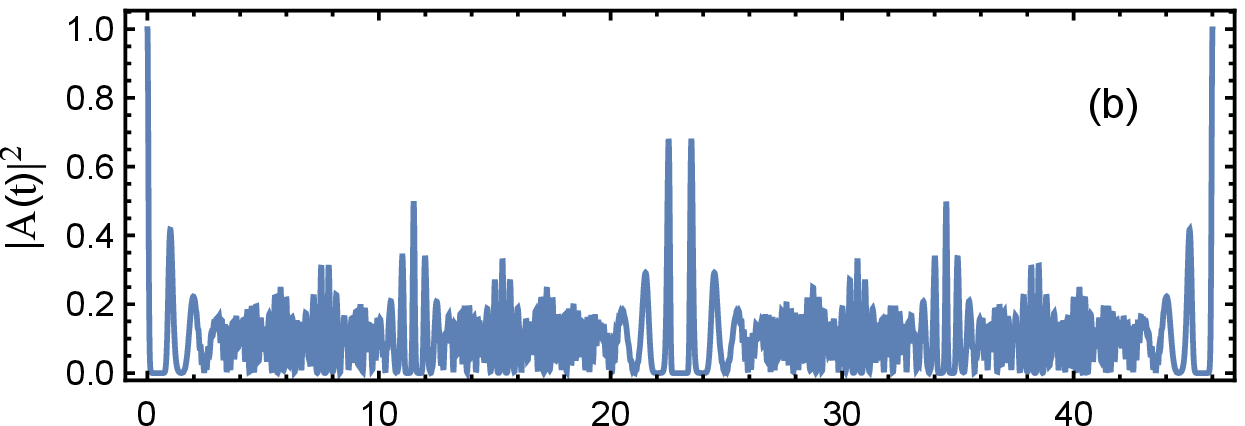}	
	\includegraphics[width=0.49\textwidth]{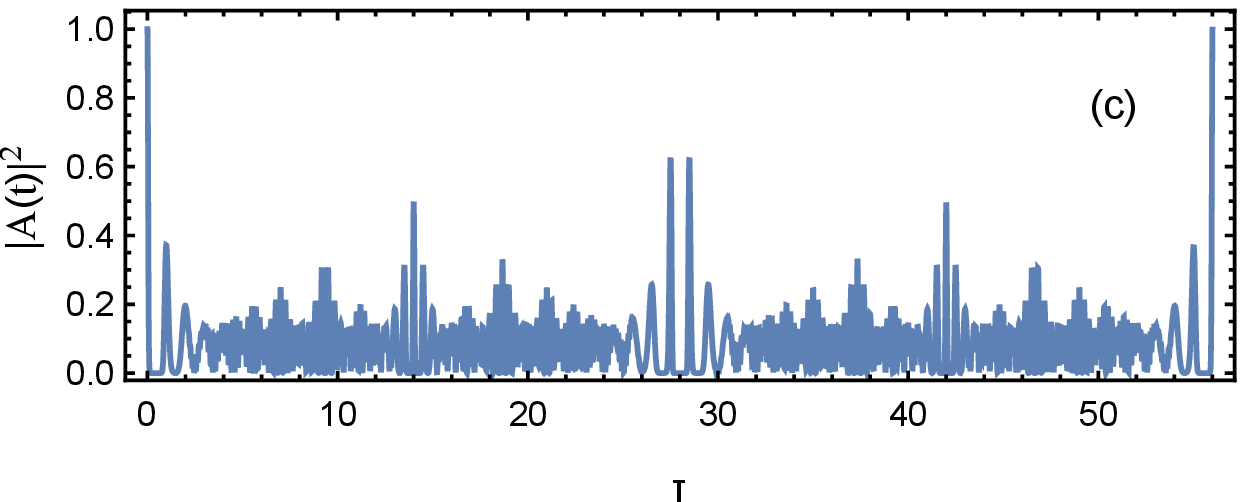}
	\includegraphics[width=0.49\textwidth]{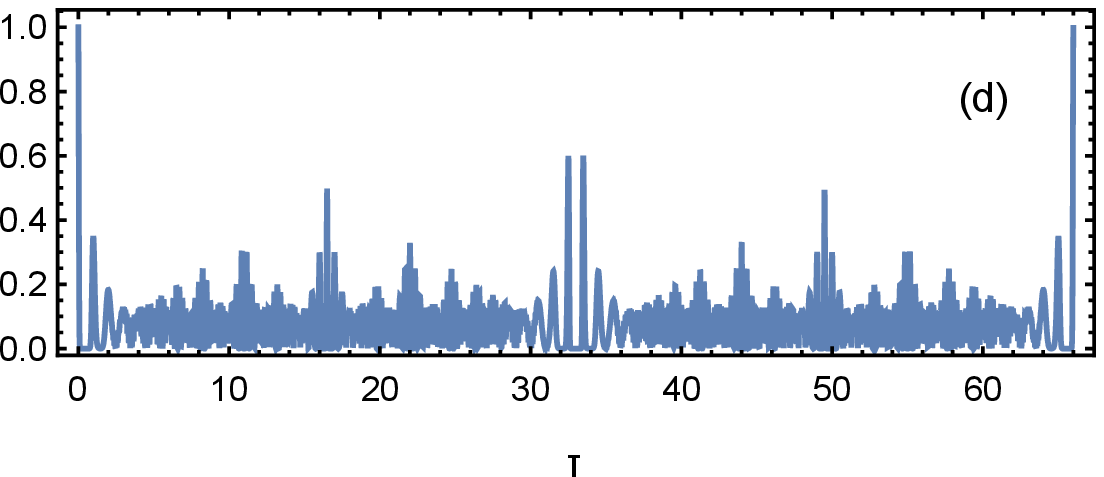}
	\caption{The modulus square of autocorrelation function $A(t)$ versus time $\tau=t/T_{cl}$ for $\upsilon=0.2$ and $(a)~n_{0}=5,~J=6.9$, $(b)~n_{0}=10,~J=15.3$, $(c)~n_{0}=15,~J=25.7,$ and $(d)~n_{0}=20,~J=38.1$.} \label{a.2}
\end{figure}

\begin{figure}
	\centering
	\includegraphics[width=0.49\textwidth]{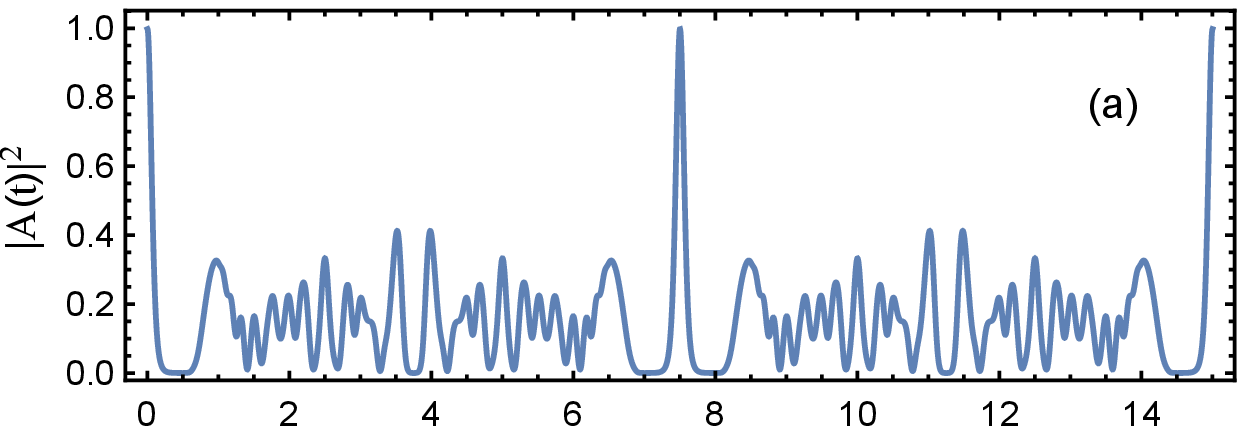}
	\includegraphics[width=0.49\textwidth]{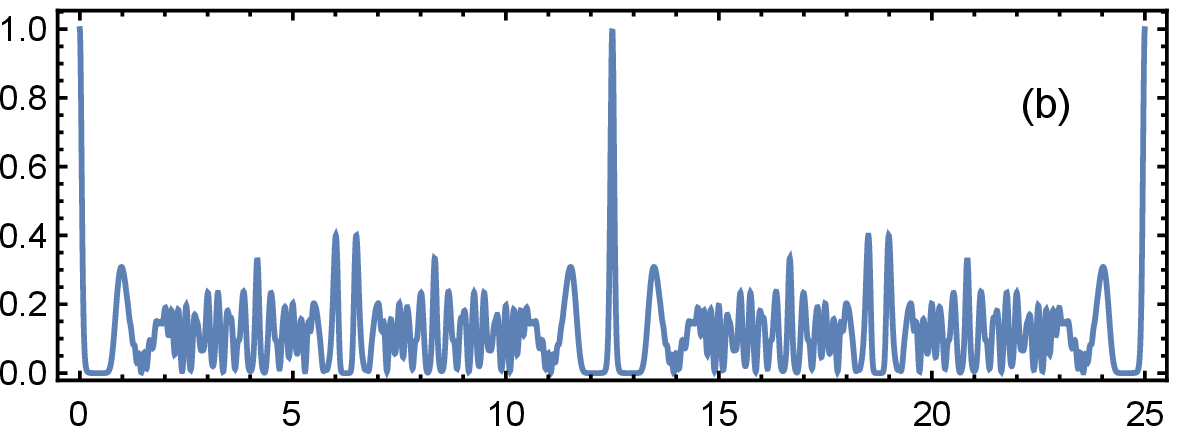}
	\includegraphics[width=0.49\textwidth]{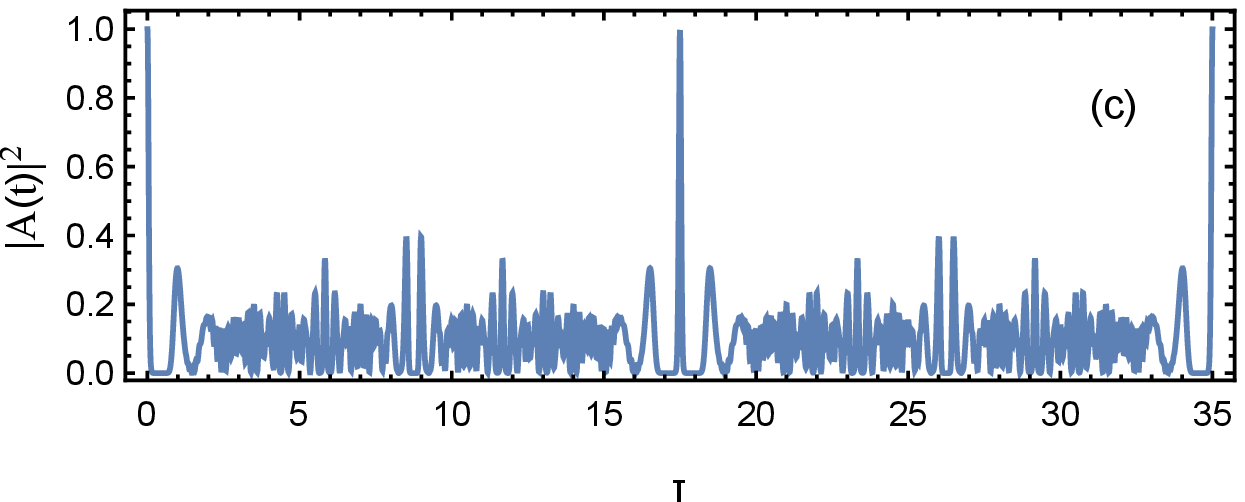}
	\includegraphics[width=0.49\textwidth]{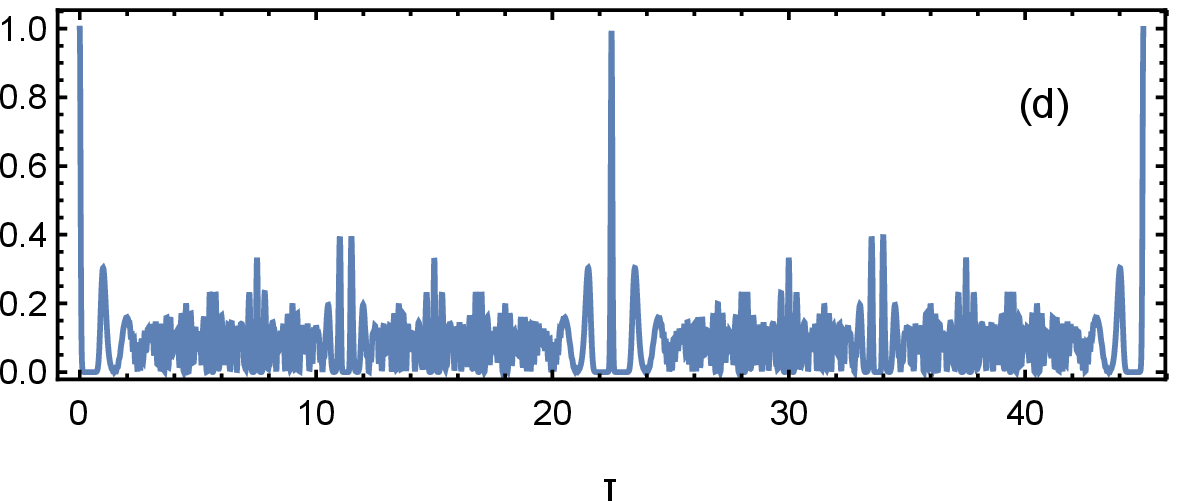}
	\caption{The modulus square of autocorrelation function $A(t)$ versus time $\tau=t/T_{cl}$ for $\upsilon=.5$ and $(a)~n_{0}=5,~J=14.3$, $(b)~n_{0}=10,~J=40.6$, $(c)~n_{0}=15,~J=79.3,$ and $(d)~n_{0}=20,~J=130.3$.} \label{a.3}
\end{figure}

\begin{figure}
	\centering
	\includegraphics[width=0.49\textwidth]{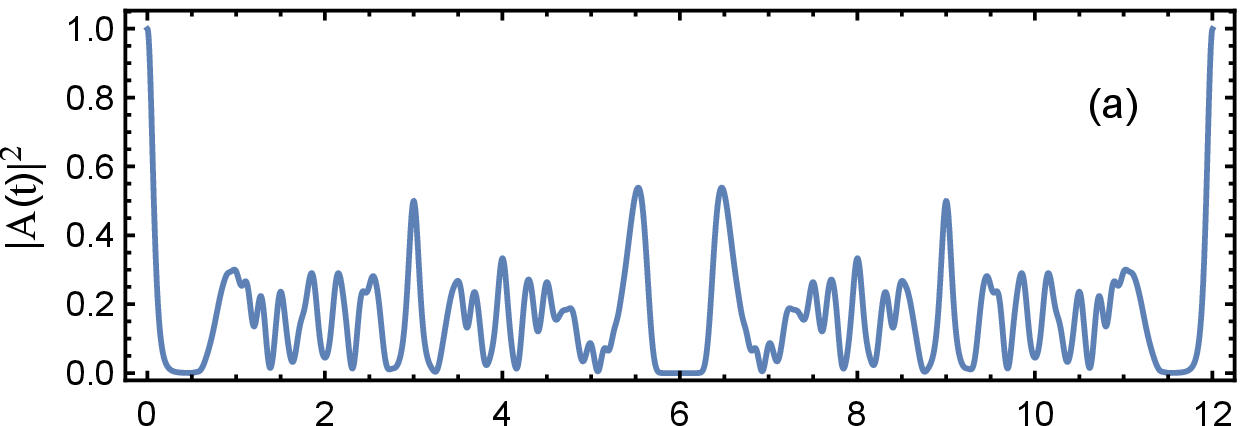}
	\includegraphics[width=0.49\textwidth]{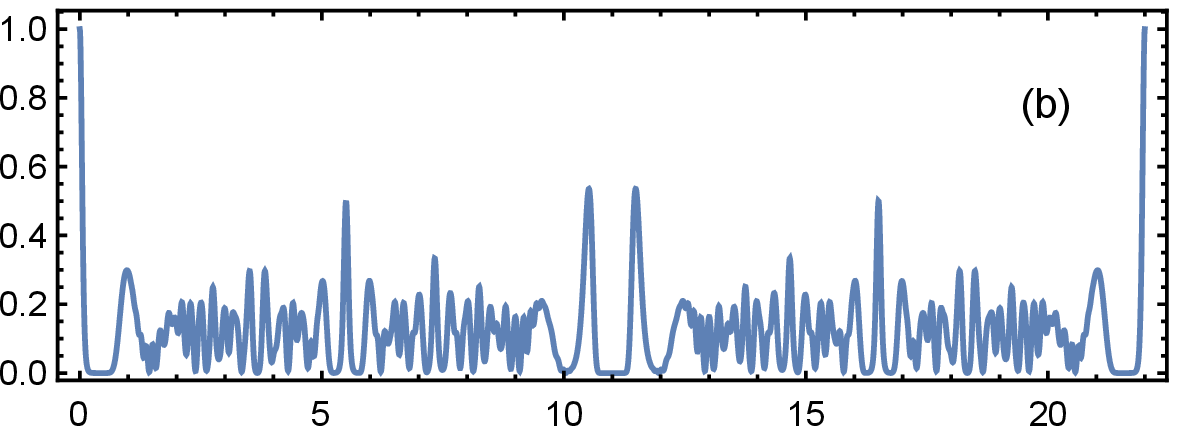}
	\includegraphics[width=0.49\textwidth]{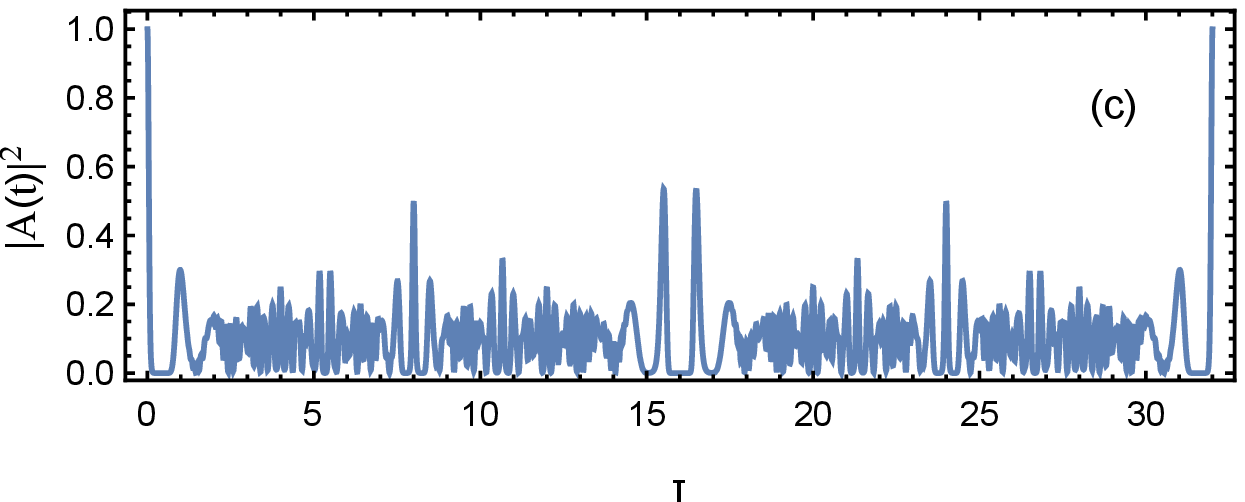}
	\includegraphics[width=0.49\textwidth]{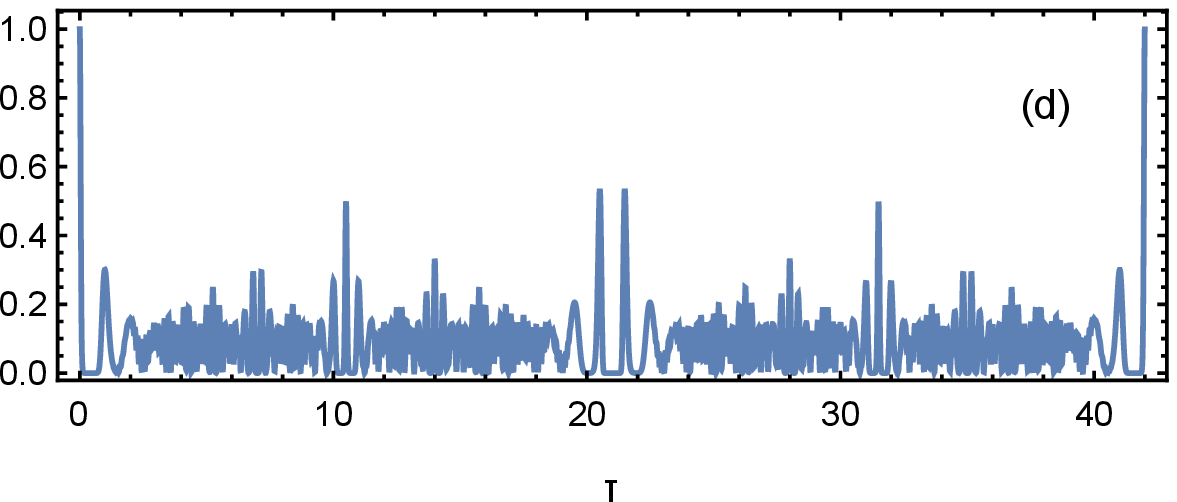}
	\caption{The modulus square of autocorrelation function $A(t)$ versus time $\tau=t/T_{cl}$ for $\upsilon=1$ and $(a)~n_{0}=5,~J=41$, $(b)~n_{0}=10,~J=131$, $(c)~n_{0}=15,~J=271,$ and $(d)~n_{0}=20,~J=459$.} \label{a.4}
\end{figure}

\indent We now investigate the temporal characteristics of the GK coherent states, obtained in Eq. (\ref{g5.8}). Using Eq.~(\ref{g5.4}) into Eq.~(\ref{Tr}), we get the classical period and quantum revival time for the coherent states of PDEM nonlinear oscillator as
\begin{eqnarray}
T_{cl}&=&\frac{2\pi}{1+\upsilon^{2}(2n_{0}+1)},
\end{eqnarray} 
and
\begin{eqnarray}
T_{rev}&=&\frac{2\pi}{\upsilon^{2}},
\end{eqnarray} 
respectively. It is important to note that the classical period $T_{cl}$ is inversely proportional to mean excitation number $n_{0}$ and in turn to parameter $J$ ( since $n_{0}$ is proportional to $J$ as shown in Fig.~(\ref{mv})). However, the quantum revival time is independent of $n_{0}$ as well as of $J$. Furthermore, we note that both, the classical period and the quantum revival time, have inverse proportion with nonlinearity parameter $\upsilon$. These facts govern the structure of quantum revivals and fractional revivals which can be explored by the autocorrelation function, given as 
\begin{equation}\label{auto'}
A(t)= \frac{1}{\mathcal{N}(J)}\sum_{n=0}^{\infty}
\bigg[\frac{\Gamma\big(2+\frac{1}{\upsilon^{2}}\big)}{n!\Gamma\big(2+\frac{1}{\upsilon^{2}}+n\big)}\bigg]
\bigg(\frac{J}{\upsilon^{2}}\bigg)^{n}e^{-i \alpha n[1+\upsilon^{2}(n+1)]t},
\end{equation}
where we have used Eqs.~(\ref{a.8})  and (\ref{g5.8}) in to general expression of autocorrelation function given in Eq.~(\ref{autocs}). In order to investigate the structure of quantum revivals and fractional revivals, we plot squared modulus of the autocorrelation function $\mid A(t)\mid^{2}$ versus time $\tau=t/T_{rev}$ for different values of coherent state parameters which are displayed in Figures (\ref{a.1})-(\ref{a.4}). 
It is obvious from these plots that the structure of fractional revivals becomes more evident as the mean excitation number $n_{0}$ increases (by increasing coherent state parameter $J$) for any fixed value of $\nu$. In contrast, higher order fractional revivals become less apparent as the nonlinearity parameter $\nu$ increases for a fixed value of $n_{0}$. 

As discussed in the begining of this section, various choices of $m(x)$ in Eq.~(\ref{vx}) lead to different potentials. For instance another particular choice of $m(x)=(1+\lambda x^{2})^{-1}$ lead to Mathews-Lakshmanan-type oscillator \cite{1,2} which has been extensively studied in various contexts \cite{ai15,aibg,aiagcs,ai14,ailo,aias}. The energy eigenvalues in this case have been obtained \cite{ai14} as 
\begin{equation}\label{a.8c2}
E_{n} = \alpha \bigg[\bigg(n+\frac{1}{2}\bigg)-\frac{\tilde{\lambda}}{2}~n(n+1)\bigg],~~~~~~~n=0,1,2,....
\end{equation}
where the parameter $\tilde{\lambda}$ can either be positive or negative and for $\tilde{\lambda}=0$ we recover the case of the standard linear harmonic oscillator with constant mass. The two cases $\tilde{\lambda}<0$ and $\tilde{\lambda}>0$ are rather different so they must be considered separately. It is important to note from Eq. (\ref{a.8c2}) that $\tilde{\lambda}<0$ lead to the same energy spectrum as given in Eq.~(\ref{a.8}) and the coherent states in this case will have the same characteristics as discussed above. On the other hand, Eq. (\ref{a.8c2}) lead to truncated spectrum for $\tilde{\lambda}>0$ and the coherent states can only be constructed for very small values of $\tilde{\lambda}$ which have been discussed in \cite{ai15}.

\subsection{Morse-like oscillator}
Let us now consider a Morse-like oscillator with position-dependent effective mass given as
\begin{equation}\label{vxm}
V(x,\alpha)=\frac{\mu^{2}}{2}m(x)[(\alpha^{2}-1)e^{2\mu x}+1]-\frac{\mu^{2}}{2}(\alpha+1),
\end{equation}
where $\mu$ is the nonlinearity parameter. For the particular mass profile  
\[m(x)=\frac{e^{-\mu x}}{2},~~\mu>0,\]
the energy spectrum in dimensionless form is obtained \cite{ailo} as 
\[E_{n}=n\mu^{2}=e_{n}.\] 
In this case the associated coherent states are given as
\begin{equation}\label{csm}
|J,\gamma\rangle=\frac{1}{\sqrt{\mathcal{N}(J)}}
\sum_{n=0}^{\infty} \frac{1}{\sqrt{n!}}\bigg(\frac{J}{\mu^{2}}\bigg)^{\frac{n}{2}}e^{-i\gamma e_{n}}|\psi_{n}\rangle,
\end{equation}
where $\rho_{n}=n! \mu^{2n}$ with $\rho_{0}=1$ and the normalization constant $\mathcal{N}(J)=e^{J\mu^{-2}}$.  

\indent It can easily be seen that the states obtained in Eq. (\ref{csm}), satisfy the Klauder's minimal set of conditions that are required for any coherent state \cite{klu.0}. The radius of convergence for the pertaining system is given as $R=\lim_{n\rightarrow \infty}(n!\mu^{2n})^{\frac{1}{n}}=\infty,$
which shows that the coherent states for the present case are defined on the entire complex plane. Moreover, the weighting distribution in this case turns out to be
\begin{equation}\label{wdm}
P_{n}=\frac{e^{-\frac{J}{\mu^{2}}}}{n!}\bigg(\frac{J}{\mu^{2}}\bigg)^{n},
\end{equation}
with corresponding mean and variance related by
\begin{equation}\label{mv3}
\langle n \rangle=\frac{J}{\mu^{2}}
~~\mbox{and}~~(\Delta n)^{2}= \frac{J}{\mu^{2}},
\end{equation}
which is the characteristic of Poisson distribution.
The time evolution of the coherent states (\ref{csm}) is given as
\begin{equation}\label{csmt}
|J,\gamma,t\rangle=\frac{1}{\sqrt{\mathcal{N}(J)}}
\sum_{n=0}^{\infty} \frac{1}{\sqrt{n!}}\bigg(\frac{J}{\mu^{2}}\bigg)^{\frac{n}{2}}e^{-i\mu^{2} (\gamma+t)n}|\psi_{n}\rangle,
\end{equation}
which shows that these coherent states are temporally stable and evolve in time with classical periodicity as in the case of linear harmonic oscillator.   

\section{Conclusion}
In this article, we have constructed coherent states for a class of nonlinear oscillators with position-dependent mass using Gazeau-Klauder formalism. Statistical properties and temporal characteristics of these constructed coherent states have been explored by means of Mendel parameter and autocorrelation function respectively. In particular we considered two kinds of nonlinear oscillator, one with linear and other with nonlinear energy spectrum with respect to quantum number. We found that the coherent states for position-dependent mass oscillators are temporally stable as long as the underlying energy spectrum is linear. Otherwise their time evolution exhibit quantum revivals and fractional revivals. We explored the structure of the fractional revivals by means of autocorrelation as a function of time for various choices of coherent states parameters. 



\begin{thebibliography}{00}
	\bibitem{e.schr} E. Schr\"{o}dinger, Naturwissenschaften \textbf{14} (1926) 664.
	
	\bibitem{G.1} R. J. Glauber, Phys. Rev. Lett. \textbf{10} (1963) 84;\\
	              R. J. Glauber, Phys. Rev. \textbf{130} (1963) 2529;\\
	              R. J. Glauber, Phys. Rev. \textbf{131} (1963) 2766.
	
%
	
	\bibitem{grj07} R. J. Glauber, \textit{Quantum theory of optical coherence: selected papers and lectures,} John Wiley $\&$ Sons, Weinheim,
		Germany ( 2007).
	
	\bibitem{gjp09} J. P. Gazeau, \textit{Coherent states in quantum physics} WILEY-VCH Verlag GmbH $\&$ Co. KGaA, Weinheim, (2009).
	
	\bibitem{a04} A. Aleixo, A. Balantekin, J. Phys. A: Math. Gen. \textbf{37} (2004) 8513.
	
	\bibitem{f} T. Fukui, N. Aizawa, Phys. lett. A \textbf{180} (1993) 308.
	
	\bibitem{amp86} A. Perelomov, \textit{Generalized Coherent States and Their Applications},
	Springer Science \& Business Media, Berlin,
	Heidelberg (2012).
	\bibitem{aagm} S. T. Ali, J. P. Antoine, J. P. Gazeau, U. Mueller, Rev. Math. Phys. \textbf{7} (1995) 1013.
	
	\bibitem{ajj} S. T. Ali, J. P. Antoine, J. P. Gazeau, \textit{Coherent states, wavelets and their generalizations}, Springer, New York (2000).
	
	\bibitem{klu.0} J. R. Klauder, J. Math. Phys. \textbf{4} (1963) 1055.
	
	\bibitem{b.g} A. Barut, L. Girardello, Commun. Math. Phys. \textbf{21} (1971) 41.
	
	\bibitem{q1}  L. C. Biedenharn, J. Phys. A {\bf22}, (1989) L873 .
	
	\bibitem{f1} V. I. Man'ko, G. Marmo, E. C. G. Sudarshan, F. Zaccaria Phys. Scr. {\bf 55} (1997) 528.
	
	\bibitem{r1} S. H. Dong, Can. J. Phys. \textbf{80} (2002) 129;\\
	S. H. Dong, M. Lozada-Cassou, Int. J. Mod. Phys. B \textbf{19} (2005) 4219;\\
	D. Popov, S. H. Dong, N. Pop, V. Sajfert, S. Şimon, Ann. Phys. \textbf{339} (2013) 122.
	                        
	
	\bibitem{Klu.1} J. R. Klauder, J. Phys. A: Math. Gen. \textbf{29} (1996) L293.
	
	\bibitem{G.K} J. P. Gazeau, J. R. Klauder, J. Phys. A: Math. Gen. \textbf{32} (1999) 123.
	
	\bibitem{spf} S. Iqbal, P. Rivi\`{e}re, F. Saif, Int. J. Theor. Phys. \textbf{49} (2010) 2540;\\
	S. Iqbal, F. Saif, J. Math. Phys. \textbf{52} (2011) 082105;\\
	S. Iqbal, F. Saif, Phys. Lett. A \textbf{376} (2012) 1531;\\
	S. Iqbal, F. Saif, J. Russ. Laser Res. \textbf{34} (2013) 77.
	

	\bibitem{pra} J. P. Antoine, J. P. Gazeau, P. Monceau, J. R. Klauder, K. A. Penson, J. Math. Phys. \textbf{42} (2001) 2349.
	
	
	\bibitem{ujp} U. Roy, J. Banerji, P. K. Panigrahi, J. Phys. A: Math. Gen. \textbf{38} (2005) 9115.
	
	\bibitem{jpb9a} J. Keeling J, V. Gurarie, Phys. Rev. Lett. {\bf 101} (2008) 033001.
	
	\bibitem{jpb10} I. L. Garanovich, S. Longhi, A. A. Sukhorukov, Y. S. Kivshar, Phys. Rep. {\bf 518} (2012) 1.
	
	\bibitem{jpb14} Z. D. Gaeta, C. R. Jr. Stroud, Phys. Rev. A {\bf 42} (1990 ) 6308.
	
	\bibitem{jpb9} P. Rivi\'{e}re P., S. Iqbal, J. M. Rost, J. Phys. B: At. Mol. Opt. Phys. \textbf{47} (2014) 124039;\\
	P. Rivi\'{e}re, S. Iqbal, J. M. Rost, J. Phys.: Conf. Ser. \textbf{194} (2009) 022015;\\
	H. Katsuki, H. Chiba, C. Meier, B. Girard, K. Ohmori, Phys. Chem. Chem. Phys. {\bf 12} (2010) 5189.
	
	\bibitem{si2006} S. Iqbal, Qurat-ul-Ann, F. Saif, Phys. Lett. A \textbf{356} 231 (2006);\\
	F. Saif, Phys. Rep. \textbf{419}, (2005) 207;\\
	M. Ayub, F. Saif, Phys. Rev. A \textbf{85} (2012) 023634.	 
	
	\bibitem{siphd} S. Iqbal, \textit{Quantum Chaos In Driven Power law Potentials:
	Generalized Coherent States To Wave Packet Evolution}, PhD Thesis, Quaid-i-Azam University, Islamabad, Pakistan (2011)

	\bibitem{buch} A. Buchleitner, D. Delande, J. Zakrzewski, Phys. Rep. \textbf{368} (2002) 409;\\
	F. Saif, Phys. Rep. \textbf{419} (2005) 207.
	
	\bibitem{rob} R. W. Robinett, Phys. Rep. \textbf{392} (2004) 1.
	
	\bibitem{is16} I. Yausaf, S. Iqbal, J. Russ. Laser Res. \textbf{37} (2016) 328.
	
	\bibitem{ap} I. S. Averbukh, N. F. Perelman, Phys. Lett. A \textbf{139} (1989) 449.
	
	\bibitem{p1} M. R. Geller, W. Kohn, Phys. Rev. Lett. \textbf{70} (1993) 3103.
	
	\bibitem{p2} G. Bastard, \textit{Wave Mechanics Applied to Semiconductor
	Heterostructures}, Les Ulis Cedex: Les \'{E}ditions de Physique, Paris (1988).
	
	\bibitem{p7} O. Von Roos, H. Mavromatis, Phys. Rev. B \textbf{31} (1985) 2294;\\
	R. A. Morrow, Phys. Rev. B \textbf{35} (1987) 8074;\\
	K. Young, Phys. Rev. B \textbf{39} (1989) 13434;\\
	G. T. Einevoll, P. C. Hemmer, J. Thomsen, Phys. Rev. B \textbf{42} (1990) 3485;\\
	R. Koc, M. Koca, G. Sahinoglu, Eur. Phys. J. B \textbf{48} (2005) 583;\\
	A. G. M. Schmidt, Phys. Lett. A \textbf{353} (2006) 459;\\
	S. H. Mazharimousavi, Phys. Rev. A \textbf{85} (2012) 034102.
	
	\bibitem{p9}A. R. Plastino, M. Casas, A. Plastino, Phys. Lett. A \textbf{281} (2001) 297.
	
	\bibitem {p11} T. Q. Dai, Y.F. Cheng, Phys. Scr. \textbf{79} (2009) 015007;\\
	S. M. Ikhdair, Eur. Phys. J. A \textbf{40} (2009) 143.
	
	\bibitem{p12}   
	A. Ganguly, L.M. Nieto, J. Phys. A \textbf{40} (2007) 7265;\\
	O. Panella,S. Biondini,A. Arda, J. Phys. A: Math. Theor. \textbf{43} (2010) 325302.
	
	\bibitem{r2} J. Yu, S. H. Dong, Phys. Lett. A \textbf{325} (2004) 194;\\
	J. Yu, S. H. Dong, G. H. Sun, Phys. Lett. A \textbf{322} (2004) 290;\\
	S. H. Dong, J. J. Pena, C. Pacheco-Garcia, J. Garcia-Ravelo, Mod. Phys. Lett. A \textbf{22} (2007) 1039.	 
	
	
	
	
	\bibitem{com}   N. Amir, S. Iqbal, J. Math. Phys. \textbf{55} (2014) 0114101.
	
	\bibitem{ai15}N. Amir, S. Iqbal, J. Math. Phys. \textbf{56} (2015) 062108.
	
	\bibitem{aibg}N. Amir, S. Iqbal, Commun. Theor. Phys. \textbf{66} (2016) 41.
	
	\bibitem{aiagcs}N. Amir, S. Iqbal, Commun. Theor. Phys. \textbf{66} (2016) 615.	
	\bibitem{von}   O. Von Roos, Phys. Rev. B \textbf{27} (1983) 7547.
	
	\bibitem{levy}  J. M. L\'{e}vy-Leblond, Phys. Rev. A \textbf{52} (1995) 1845.
	
	\bibitem{ai14} N. Amir, S. Iqbal, Commun. Theor. Phys. \textbf{62} (2014) 790.
	\bibitem{ailo}N. Amir, S. Iqbal, Europhys. Lett. \textbf{111} (2015) 20005.
	
	\bibitem{aias}N. Amir, S. Iqbal, J. Math. Phys. \textbf{57} (2016) 062105.
	
	\bibitem{1}    P. M. Mathews, M. Lakshmanan, Quart. Appl. Math. \textbf{32} (1974) 215.
	
	\bibitem{2}   M. Lakshmanan and S. Rajaseekar, \textit{Nonlinear Dynamics:
Integrability, Chaos and Patterns}, Springer Science
$\&$ Business Media, Berlin, Heidelberg (2012).
	
	\bibitem{3}   R. Delbourgo, A. Salam, J. Strathdee, Phys. Rev. \textbf{187} (1969) 1999.
	

	\bibitem{6}    J. F. Cari\~{n}ena, M. F. Ra\~{n}ada, M. Santander, Rep. Math. Phys. \textbf{54} (2004) 285.
	
	\bibitem{7}    J. F. Cari\~{n}ena, M. F. Ra\~{n}ada, M. Santander, Ann. Phys. (N.Y.) \textbf{322} (2007) 2249.
	\bibitem{m09} B. Midya, B. Roy, A. Biswas, Phys. Scr. \textbf{79} (2009) 065003.
	
	\bibitem{b09} A. Biswas, B. Roy, Mod. Phys. Lett. A \textbf{24} (2009) 1343.
	
	\bibitem{r10} V. C. Ruby, M. Senthilvelan, J. Math. Phys. \textbf{51} (2010) 052106.
	
	\bibitem{sco} S. C. y Cruz, O. Rosas-Ortiz, Int. J. Theor. Phys. \textbf{50} (2011) 2201.
	
	\bibitem{sg12} S. Ghosh, J. Math. Phys. \textbf{53} (2012) 062104.
	
	\bibitem{sam} S. A. Yahiaoui, M. Bentaiba, J. Phys. A: Math. Theor. \textbf{47} (2014) 025301.
	
	
	\bibitem{ags} A. G. Schmidt, Phys. Lett. A \textbf{353} (2006) 459.
	

	
	\bibitem{cooper1995supersymmetry} F. Cooper, A. Khare, U. Sukhatme, Phys. Rep. \textbf{251} (1995) 267.
	
	\bibitem{cooper2001supersymmetry} F. Cooper, A. Khare, U.P. Sukhatme, \textit{Supersymmetry
	in Quantum Mechanics}, World Scientific, Singapore
	(2001).
	
	\bibitem{milanovic1999generation} V. Milanovic, Z. Ikonic, J. Phys. A: Math. Gen. \textbf{32} (1999) 7001.
	
	\bibitem{plastino1999supersymmetric} A. Plastino, A. Rigo, M. Casas, F. Garcias, A. Plastino, Phys. Rev. A \textbf{60} (1999) 4318.
	
	\bibitem{gonul2002supersymmetric} B. Gonul,  B. Gonul, D. Tutcu, O. Ozer, Mod. Phys. Lett. A \textbf{17} (2002) 2057.
	
	\bibitem{gendenshtein1983derivation} L. Gendenshtein, JETP. Lett. \textbf{38} (1983) 356.
		
	\bibitem{b} A. B. Balantekin, Phys. Rev. A \textbf{57} (1998) 4188.
		
	\bibitem{bagchi2005deformed} B. Bagchi, A. Banerjee, C. Quesne, V. Tkachuk, J. Phys. A: Math. Gen. \textbf{38} (2005) 2929.
	
	\bibitem{samani2003shape} K. Samani, F. Loran, arXiv preprint quant-ph/0302191 (2003).
	
	\bibitem{ganguly2007shape} A. Ganguly, L. Nieto, J. Phys. A: Math. Theor. \textbf{40} (2007) 7265.
	
	\bibitem{tezcan2007exact} C. Tezcan, R. Sever, J. Math. Chem. \textbf{42} (2007) 387.
	
	\bibitem{mustafa2006d} O. Mustafa, S. H. Mazharimousavi, J. Phys. A: Math. Gen. \textbf{39} (2006) 10537.
	
	\bibitem{quesne2009point} C. Quesne, SIGMA \textbf{5} (2009) 046.
	
	\bibitem{kamran1990lie} N. Kamran, P.J. Olver, J. Math. Anal. Appl. \textbf{145} (1990) 342.
	
	\bibitem{roy2005lie} B. Roy, Europhys. Lett. \textbf{72} (2005) 1.
	
	\bibitem{chetouani1995green} L. Chetouani, L. Dekar, T. F. Hammann, Phys. Rev. A \textbf{52} (1995) 82.
	
	\bibitem{mandal2000path} B. P. Mandal, Int. J. Mod. Phys. A \textbf{15} (2000) 1225.
	
	
	\bibitem{mandel}  L. Mandel, Opt. Lett., \textbf{4} (1979) 205;\\
	L. Mandel, E. Wolf, \textit{Optical Coherence and Quantum Optics}, Cambrige University Press, Cambrige, (1995).
	
	\bibitem{rkam} 	A. Mathai, R.K. Saxena, \textit{Generalized Hypergeometric
	Functions with Applications in Statistics and Physical
	Sciences}, Springer, Berlin, Heidelberg (1973).
	
\end{thebibliography}
\end{document}